\documentclass[letter]{aa} % for the letters 
\usepackage{graphicx}
\usepackage{txfonts}
\usepackage{amsmath}
\usepackage{siunitx}
\usepackage{xcolor}
\usepackage{longtable} %for a long table spanning multiple pages
\usepackage[online]{threeparttablex} %to add notes under a table
\usepackage{array} %to solve the size problem of the longtable

\usepackage{hyperref}
\hypersetup{colorlinks=true,linkcolor=blue, citecolor=blue,  linktocpage}

\usepackage{natbib}
\bibpunct{(}{)}{;}{a}{}{,} % to follow the A&A style
\usepackage{aas_macros}
\usepackage{adjustbox} %to crop pgf
\usepackage{printlen} %size of column
\usepackage{supertabular} % table over 2 columns
\usepackage{booktabs} %same but other method
\usepackage{todonotes}

\usepackage{diagbox}

\usepackage{subcaption}

\newcommand{\comment}[1]{}

\begin{document} 

 \title{Discovery of the first olivine-dominated A-type asteroid family}

   \author{M. Galinier\inst{1}
          \and
          M.~Delbo \inst{1,2}
          \and
          C. Avdellidou\inst{2,1}
          \and
          L. Galluccio\inst{1}}

\institute{Universit\'e C\^ote d'Azur, CNRS--Lagrange, Observatoire de la C\^ote d'Azur, CS 34229 -- F 06304 NICE Cedex 4, France\\
\email{marjorie.galinier@oca.eu}
\and 
University of Leicester, School of Physics and Astronomy, University Road, LE1 7RH, Leicester, UK
}

\date{Received date, year; accepted date, year}

% \abstract{}{}{}{}{} 
% 5 {} token are mandatory 
 \abstract
 {
The classical theory of differentiation states that due to the heat generated by the decay of radioactive elements, some asteroids form an iron core, an olivine-rich mantle, and a crust. The collisional breakup of these differentiated bodies is expected to lead to exposed mantle fragments, creating families of newly-formed asteroids. Among these new objects, some are expected to show an olivine-rich composition in spectroscopic observations. However, several years of spectrophotometric surveys have led to the conclusion that olivine-rich asteroids are rare in the asteroid main belt, and no significant concentration of olivine-rich bodies in any asteroid family has been detected to date. Using ESA’s Gaia DR3 reflectance spectra, we show that the family (36256) 1999~XT17 presents a prominence of objects that are likely to present an olivine-rich composition (A-type spectroscopic class). If S-complex asteroids as the second most prominent spectroscopic class in the family are real family members, then arguably the 1999~XT17 family has originated from the break-up of a partially differentiated parent body. Alternatively, if the S-complex asteroids are interlopers, then the 1999~XT17 family could have originated from the breakup of an olivine-rich body. This body could have been part of the mantle of a differentiated planetesimal, which may have broken up in a different region of the Solar System, and one of its fragments (i.e. the parent body of the 1999~XT17 family) could have been dynamically implanted in the main belt.
}

\maketitle
%
%-------------------------------------------------------------------

\section{Introduction}
\label{intro}

Early-generation planetesimals are bodies that formed $\lesssim$2~Myr after the formation of our Solar System. They are believed to have gone through a process of differentiation \citep{sramek2012,kruijer2014}, which causes a body to organise into layers of different densities and composition, due to the heat generated by the decay of radioactive elements, such as $^{26}$Al. In the classical scenario of asteroid differentiation, these layers comprise a basaltic crust, an olivine-rich mantle, and an iron core \citep{elkins-tanton2017}. 

Asteroid differentiation is corroborated by the existence of meteorites originating from the different aforementioned layers, such as iron meteorites \citep{cloutis1990}, mesosiderites \citep{hewins1983,greenwood2006}, brachinites and pallasites \citep{cruikshank1984}, and andesites \citep{barrat2021}, as well as the howardite-eucrite-diogenite (HED) meteorites \citep{mccord1970}. However, apart from few clear cases such as the asteroid (4) Vesta \citep{russell2012}, the detection of differentiated asteroids and the assessment of their abundance in the main belt remains a challenge.

One way to detect a potentially differentiated asteroid is to discover a very-high-density object, indicating the presence of an iron core. Although large improvements were made in the last two decades in the measurement of asteroid sizes \footnote{see references at \url{mp3c.oca.eu}} and volumes \citep[e.g.][]{durech2010,durech2011,vernazza2021}, mass estimations \citep{fienga2020} are still limited and may suffer from large uncertainties. This issue leads to a reliable density estimation for a relatively small number of objects in the main belt and very-high-density objects detected may be due to errors \citep{carry2012}. A related issue is that large collisions may have induced voids in a body, causing a potential high density to be camouflaged and thus hiding hints of differentiation.

Another way to discover differentiated objects is to study collisional families of asteroids. In principle, if a large impact event shatters a differentiated asteroid, the resulting family of newly born smaller asteroids should contain fragments sampling the layers of the parent body. Hence, a variety of compositions should be detectable within the collisional family. However, a catastrophic disruption of the parent body is needed to reveal such layers. Indeed, cratering events resulting from impacts may only produce ejecta from the parent body's crust. One example is the Vesta family, which resulted from the two impacts that formed the large basins Rheasilvia and Veneneia on (4) Vesta \citep{marchi2012}. Remote sensing data from NASA’s Dawn space mission \citep{russell2012}, ground-based spectroscopy, and geometric visible albedo data, have revealed a remarkable link between the crust of (4) Vesta, its family members, and the HED meteorites. The basaltic composition of Vesta family members indicates that Vesta is a differentiated asteroid, but, despite its large extension, Vesta family members do not clearly show mantle and core material compositions. 

Spectroscopic searches for traces of differentiated material in the main belt have led to the discovery of basaltic \citep{mccord1970}, olivine-rich \citep{demeo2019}, and potentially metal-rich asteroids \citep{fornasier2010,harris&drube2014}. However, different spectroscopic surveys shed light on the scarcity of olivine-rich asteroids compared to other asteroid types in the main belt. This lack of mantle-like asteroids is known as the `missing mantle problem' \citep{burbine1996}.
Olivine-rich asteroids show a moderately high albedo, and a spectrum characterised by extremely reddish slopes short-wards of 0.7~\SI{}{\micro\meter}, a strong absorption feature centred at about 1.05~\SI{}{\micro\meter}, and an absent or weak absorption feature at 2~\SI{}{\micro\meter}. These objects are classified as A-types in most taxonomic schemes \citep{bus2002tax,demeo2009,Mahlke2022}. 

Using spectroscopic observations in the near-infrared (NIR) wavelength range, \cite{demeo2019} found that the olivine-rich asteroid abundance in the main belt is as low as 0.16\%, for asteroids with diameters $D>$2~km. In addition, olivine-rich asteroids are found to be evenly distributed throughout the main belt, showing no statistically significant concentration in any asteroid family \citep[see ][]{sanchez2014,demeo2019}. Several theories have been developed to explain the aforementioned observations: i) A-type asteroids are highly sensitive to space-weathering phenomena, and they can no longer be detected \citep{Popescu2018_olivine}; ii) olivine-rich objects were shattered into smithereens by collisions and they have become too small to be detected by current observational techniques \citep[][]{sanchez2014,demeo2019}; iii) differentiated planetesimals formed early and accreted into planets in their majority, leaving behind very few differentiated bodies to be implanted in the main belt as intact bodies or fragments; iv) the olivine is masked by some material of another type on the asteroids \citep{demeo2019}; or v) current theories about planetesimal differentiation and formation of thick olivine mantles are incorrect \citep{demeo2019}.

In this work, we use the 60,518 Solar System objects (SSO) reflectance spectra included in the Gaia Data Release 3 (DR3) \citep{galluccio2022} to study the abundance of olivine-rich asteroids in families. In particular, by using the Gaia DR3 and ground-based asteroid spectra, we test the hypothesis that there is no significant concentration of A-type asteroids in any asteroid family in the main belt. This work is focused primarily on the catalogue of asteroid family membership from \cite{nesvorny2015}. In the following, we present the dataset used to select the family of asteroids most likely to contain a high abundance of A-types. Then, we characterise this family using the Gaia DR3, literature data, and a newly acquired reflectance spectrum. We discuss our findings in the final section.

%--------------------------------------------------------------------
\section{Family selection}
\label{family_selection}

%The data sets used for this project include asteroid proper orbital elements and reflectance spectra.
%Say that if an asteroid was classified from NIR spectroscopy, its classification was considered over others. Phot < Spec VIS < Spec NIR.
Using the families and family memberships from \cite{nesvorny2015}, and the asteroid classification from several published sources (see Appendix~\ref{app:lit_ref}), we calculated the proportion of potential A-type asteroids in each asteroid family.
To do so, we calculated the ratio between the number of objects classified at least once as A-type in different taxonomies, or as Ad-type from \citet{Popescu_2018_classification}, and the total number of asteroids with any other spectral type in each family. Then, we considered only the families having a potential A-type abundance of above 10\%. These families are (10811) Lau (FIN 619), (36256) 1999~XT17 (FIN 629), and (7468) Anfimov (FIN 635), which were found to contain 13, 23 and 28\% of potential A-type members, respectively. We looked into the Gaia DR3 data set \citep[][]{galluccio2022} for spectra of these families' members. We found that these families have 1, 15, and 7 asteroids with a spectrum in the DR3 dataset, respectively. Since only asteroid (10811) Lau itself shows a spectrum in this data set, we did not consider the Lau family in this study. For the two other families, we studied the average signal-to-noise ratio (S/N) calculated by \citep{galluccio2022}, of the DR3 spectra of the members of each family. We found that the Anfimov family shows a mean S/N of 23, with asteroids (7468) Anfimov and (18853) 1999~RO92 having the spectra with the maximum S/N, respectively of 47.41 and 32.21. These two objects had already been characterised based on spectroscopy \citep{demeo2019,Mahlke2022}, with the other members showing a low S/N; thus, we chose to focus on the family (36256) 1999~XT17 for the purposes of this study. This family shows an average S/N of 40, with nine objects having a S/N$>21$, which is considered a threshold for reliable spectral classification in the visible (VIS) wavelength range \citep{delbo2012}. This family appears to be the most suited for our search for potential A-type members.
The 1999~XT17 family is located in the so-called `pristine zone' of the main belt \citep{Broz2013_pristine}. This zone extends between 2.825 and 2.955~au and is bounded by the 5:2 and 7:3 mean motion resonances (MMRs) with Jupiter. The 1999~XT17 family contains 58 members and it has been reported to be a S-complex family by \cite{nesvorny2015}.
%A table of the members of this family having been characterised in the literature is presented in Table \ref{tab:36256_lit} in Appendix.

%--------------------------------------------------------------------
\section{Characterisation of the (36256) 1999~XT17 family}
\label{charac}

%1) gaia data classification
We searched in the literature for additional spectral data for the 15 members of the 1999~XT17 family having a DR3 reflectance spectrum (Table \ref{tab:1}). We found a NIR spectrum for asteroid 36256, from observations of \cite{demeo2019}. For asteroids 25356 and 40671, we used NIR observations from the MOVIS catalogue \citep{Popescu_2018_classification}; and for asteroids 36256 and 99004, we found visible-light spectrophotometry data from the Sloan Digital Sky Survey (SDSS) \citep{demeo2013}. We also performed NIR spectroscopic observations at the NASA Infrared Telescope Facility (IRTF) of the asteroid 33763 (Appendix~\ref{app:NIR_obs}).

%% data set : DeMeoCarry2013, DR3, NIR
%normalise
%do a Chi2 between template and the VIS and NIR parts of the spectra independently
% Chi2 = Chi2_Gaia + Chi2_NIR + Chi2_SDSS
%reduced Chi2 to normalise

%say that for the Gaia data, the first and last two bands were not taken into account, and only the bands flagged by a 0 were considered. How were the inependent data set treated ? Were the Bus-DeMeo templates sampled like the different data sets to be able to do the matching, or were the spectra of the different data sets and the templates resampled using a cubic smoothing spline function ? Or was every data set sampled like the Bus-DeMeo templates ?

Using the aforementioned spectra, we followed the method of \citet{Avdellidou2022} to classify each 1999~XT17 family member in the Bus-DeMeo taxonomy \citep{demeo2009}, as this scheme extends to the NIR. For each asteroid, a reflectance curve-matching was performed between a given spectrum and the spectral class templates of \citet{demeo2009}, by minimising the value of a reduced $\chi^{2}$ figure of merit. For example, asteroid 36256 has a VIS Gaia DR3 and a NIR spectrum, as well as SDSS spectrophotometry. Therefore, three independent $\chi^{2}$ were calculated for this object, between the three independent data sets and the Bus-DeMeo templates. The DR3 spectra were considered from 0.45 to 0.90~\SI{}{\micro\meter} only. Then, the reduced $\chi^{2}$ obtained for each data set were summed, to obtain a global value of this figure of merit. The two classes with the lowest value of the global $\chi^{2}$ represent the best two classes for a given asteroid \citep[details in Appendix B of][]{Avdellidou2022}.
We visually inspected each spectrum to check the best matching classes given by the automatic procedure. The best two classes obtained are presented in Table \ref{tab:1}.
Our algorithm classified 12 asteroids out of 15 to be A-types, as first or second best match. The remaining three objects, namely, asteroids 34902, 83124, and 88057, were classified as S-complex objects as the first or second best match.

%2) literature info
In the literature, 7 objects out of the 15 that have a spectrum in the DR3 dataset have been classified. Among them, only asteroid 36256 has been characterised by \cite{demeo2019} using NIR spectroscopy, and is found A-type by \cite{demeo2019} and \cite{Mahlke2022} in their respective taxonomic schemes. Asteroids 25356 and 40671 are found to be Ad types by \cite{Popescu_2018_classification}, from MOVIS NIR colours; and asteroids 27565, 34902, 66676, and 99004 are found to be part of the S-complex from spectrophotometric observations \citep{carvano2010,demeo2013,SergeyevCarry2021,Sergeyev2022}. Fig.\ref{fig:all_spectra} shows the Gaia DR3 spectra of the 15 family members, along with supplementary data when available, including the NIR spectrum of 33763 that we obtained. The latter confirms that 33763 is an A-type.

Among the 43 members of the 1999~XT17 family that do not have a DR3 spectrum, 21 are classified in the literature based on spectrophotometric data in the VIS or NIR wavelength range. Among these, asteroid 76627 is classified as Ad from NIR colours \citep{Popescu_2018_classification}, while 58777 and 201232 are classified as A \citep[][]{carvano2010,Sergeyev2022}. Asteroid 254896 is found A by \cite{carvano2010} and S by \cite{Sergeyev2022}. Seven asteroids (40427, 204222, 222898, 223304, 271103, 285502, and 362895) are classified as S-complex members \citep{carvano2010,demeo2013,SergeyevCarry2021,Sergeyev2022}, and asteroid 62676 is classified as S-type by \cite{Sergeyev2022} and X-type by \cite{carvano2010}. The remaining objects are classified as L, K, C, D, V or X-types (not considering the objects classified as U here).

In total, 36 members of the 1999~XT17 family have a classification. Among them, 16 are classified as A-type, which makes a proportion of potential A-types of 44.4\%. Among these, four asteroids are confirmed A-types from the combination of DR3 and NIR data. The second most abundant class in this family is the S-complex, with a proportion of potential S-complex asteroids of 30.5\%. 

\begin{figure}[h]
\includegraphics[width=0.5\textwidth]{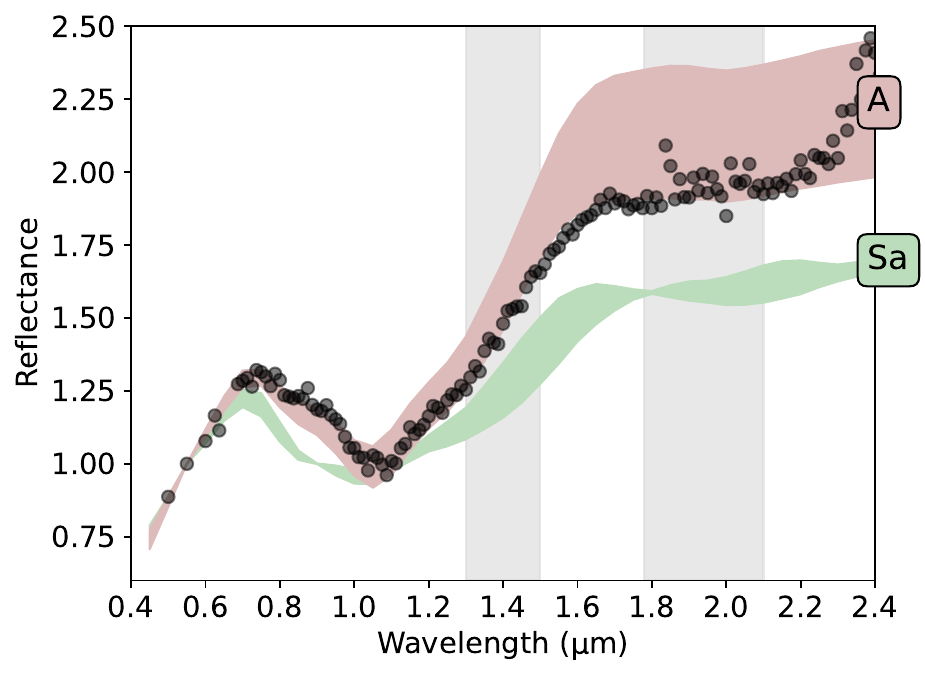} 
\caption{Average spectrum of the (36256) 1999~XT17 family members classified here as A-types, along with the A (red area) and Sa (green area) Bus-DeMeo templates.} %The DR3 spectra were considered in the wavelength range from 0.45 to 0.95~\SI{}{\micro\meter}.}
\label{fig:average}
\end{figure}

%3) average spectrum
For each asteroid with a Gaia DR3 spectrum, we combined the independent data sets used in the classification to produce a single spectrum per object. Then, following the procedure of \cite{Avdellidou2022}, we averaged the single spectra of the 12 asteroids that we classified as A-types, to produce a family averaged reflectance spectrum (Fig.\ref{fig:average}) that shows an excellent match to the A-type template of the Bus-DeMeo taxonomy. Using literature data\footnote{\url{mp3c.oca.eu}}, the uncertainty-weighted average of the geometric visible albedo ($p_V$) of the potential A-type members of the family was calculated to be 0.22, with an rms value of 0.07. %\citep[][reports an albedo of 0.21 for this family]{nesvorny2015}. Nesvorny uses a sub-sample of available albedos from the literature. Not sure you need to mention this here.
This value is within the range of the A-types' $p_V$ \citep{Mahlke2022}.

\begin{table}[!ht]
\centering
\caption{Members of the (36256) 1999~XT17 family with a Gaia DR3 spectrum.}
\begin{tabular}{r r r r r c}
\hline
Asteroid & D & $p_V$ & Avg & H & spectral  \\
Number &  (km) 		&		& S/N & & class \\
\hline
15610  	& 	5.84	&	0.30	&   35.32 &  13.3 & A, L  \\
16789  &	6.62	&	0.18	&	  38.25  & 13.5   & A, Sv  \\
20975  &	5.24	&	0.28	&	  36.01  & 13.6   & A, Sv  \\
25356  &	-	&	-	&             66.49  & 12.8   & A, Sa  \\
27565  	&	5.92	&	0.19	& 20.74   &13.6    & A, L  \\
33763  	&	-	&	-	&	       48.07  & 13.4   & A, Sa  \\
34902  	&	4.15	&	0.15	&	14.7  & 14.4    & Sq, Sr  \\
36256  	&	10.21  &	0.19	&   66.56 & 12.4   & A, Sa  \\
40671  &	-	&	-	&	        28.09 &  13.6   & A, Sa  \\
57276  	&	7.20	&	0.31	&   128.7 & 12.9   & A, Sv  \\
66676  	&	6.55	&	0.22	&  52.28  & 13.3    & L, A  \\
83124  &	-	&	-	&	        14.88 &  14.7   & S, Sr  \\
88057  	&	-	&	-	&	       20.75  & 14.5    & K, S  \\
99004  	&	3.83	&	0.10	&  13.93  & 14.6   & L, A  \\
140349 	&	3.69	&	0.20	&  13.86 &   14.7  & Sv, A  \\
\hline
 \label{tab:1}
 \end{tabular}
\tablefoot{Here is given the diameter D of family members, their geometric visible albedo $p_V$, the average signal-to-noise ratio of their DR3 spectrum S/N, their H magnitude, and the two best spectral classes given by our classification algorithm. The physical properties data were obtained from \url{mp3c.oca.eu}.}
\end{table}

%%%% proposition, to be modified
To gain insights into on the composition of these family members, we compared the two VNIR spectra of asteroids 36256 and 33763 with those of the meteorites and samples available in the Relab database (see Appendix~\ref{app:comp_met} for details). As the best spectral analogues, we obtained mostly olivine assemblages and olivine-rich meteorites for both asteroids, confirming previous findings on other A-types \citep{demeo2022}. Calculating the equivalent geometric visible albedo \citep{Beck2021} of the Relab meteorite samples and comparing it with the family's mean $p_V$, following the method developed in \cite{Avdellidou2022}, we found that the CK6 meteorite LEW87009, the R-chondrites Rumuruti and LAP04840, the R4 chondrite MIL07440, the olivine of the shergottite ALHA77005, and the brachinites NWA4882 and ALH84025, are the best matches for asteroids 36256 and 33763. However, the shergottite ALHA77005 is identified as a martian meteorite; thus, we conclude it does not come from the outer-main belt 1999~XT17 family.

%%% grouping of 2 paragraphs after having moved Fig.2 to appendix
To gain further information on the characterised family members, we studied their positions in the proper orbital element space, and in the $H$ versus proper semimajor axis $a$ space. In the same spaces, we display the neighbouring (221) Eos family as defined by \cite{nesvorny2015}. In the $H$ versus $a$ space (panel c of Fig.\ref{fig:Vshape}), the 1999~XT17 family members with $H\lesssim$~14 lay outside the V-shape of the Eos family. Interestingly, we classified them as A-types and they have a S/N$>21$. In the $e$ versus $a$ space (panel a of  Fig.\ref{fig:Vshape}), we see that the members of the 1999~XT17 family have a global proper eccentricity higher than the majority of Eos family members. In the $\sin i$ versus $a$ space (panel b of Fig.\ref{fig:Vshape}), however, members of the 1999~XT17 family show similar proper inclination values to the Eos family members.

\section{Discussion}
\label{discussion}

First, we investigated the possibility that our A-type classification based on Gaia data is affected by an issue in the DR3, which tends to overestimate the reflectance values mostly in the reddest bands \citep{galinier2023,galluccio2022}. Due to this reddening, a non-A-type asteroid could potentially be classified as such, since this class shows a red reflectance in the VIS wavelength range (see Table \ref{tab:confusion_matrix} for the confusion matrix on the A-type classification). We calculated a false positive rate (asteroids that are not A-type in the literature, but that are classified A-type by our method) smaller than 2.5\%. Therefore, we concluded that it is unlikely that our classification is affected by the aforementioned DR3 reddening. Moreover, the true positive rate (asteroids that are A-types in the literature correctly classified here) is higher than 71\% (Appendix~\ref{app:algo}). This adds confidence to our classification methods for A-types. However, we suggest using spectral matching between the DR3 and Bus-DeMeo templates with caution overall, since we investigated only the A-type case here.

In addition, we found that four asteroids among the six false negatives (asteroids that are A-type in the literature not classified as A here) in the A-type confusion matrix (Appendix~\ref{app:algo}) have a S/N$<21$. This is below the threshold commonly accepted for a reliable classification \citep{delbo2012}. Thus, it is not impossible that the three objects of the 1999~XT17 family that we did not classify as A-types could, in fact, turn out to be real A-types. Similarly, the three objects with a low S/N that we classified as A-types have to be taken with more caution.

Assuming that our classification and that the literature classes assigned from spectrophotometry are correct, then their combination indicates that the dynamical 1999~XT17 family is mainly composed of A-type objects, with a secondary S-complex class. In the following, we explore the possibility that some of these S-complex objects are interlopers of the 1999~XT17 family.

First, we studied the S-complex families in the pristine zone, assuming that they could be the source of this secondary compositional class. According to \cite{nesvorny2015}, there are other five families in the pristine zone that show a S-complex composition: (158) Koronis, (832) Seraphina, (918) Itha, (10811) Lau, and (15477) 1999~CG1. However, none of these families is in close vicinity to the 1999~XT17 family, making them an unlikely origin for the contamination.

While the 1999~XT17 family resides in a relatively empty zone compared to other regions of the main belt, it was shown that this `pristine zone' is contaminated by the Eos family \citep[][]{tsirvoulis2018,BrozMorby2019}. The 1999~XT17 family is located just inwards of the 7:3 MMR with Jupiter, in the same region as Eos family members that drifted away from the centre of the family by the action of non-gravitational forces \citep{nesvorny2015}. These `Eos family fugitives' crossed the 7:3 MMR fast enough so that their eccentricity did not reach planet crossing values. This resonance crossing can lead to a discontinuity in eccentricity between Eos fugitives and the Eos core family members \citep{nesvorny2015}. On the other hand, this resonance crossing affects little the proper inclination of the objects.

Interestingly, every 1999~XT17 family member belonging to the secondary S-complex lays inside the Eos family V-shape, and presents similar inclination values and a discontinuity in eccentricity compared to the Eos family core members (see panels a and b of Fig.\ref{fig:Vshape}, and \cite{BrozMorby2019}). These objects could therefore be Eos family fugitives that were merged with asteroids of the 1999~XT17 family by the hierarchical clustering method used for family identification \citep{nesvorny2015}. Spectroscopically, Eos is mostly composed of K-type asteroids \citep{BrozMorby2019}. However, the distinction between S and K-types is non trivial in the visible for the Eos family members \citep{Vok2006}. We conclude that it is not impossible that the S-complex asteroids of the 1999~XT17 family are part of the Eos fugitives.

On the other hand, the cluster of asteroids laying outside the Eos family V-shape contains ten asteroids, nine of which are classified as A-types from their DR3 spectra with a S/N$>$21. Asteroids 36256 and 33763 are confirmed A-types from NIR spectroscopy, while 25356 and 40671 are confirmed A-types from the combination of their Gaia DR3 spectrum and their NIR colours \citep{Popescu_2018_classification}. Moreover, the link we established between asteroids 36256 and 33763 and olivine-rich samples and meteorites from the Relab database, confirms the likely high content in olivine of these bodies. Since these objects are out of the V-shape of Eos, they are less likely to be part of Eos family fugitives. We therefore argue that the A-type members of the 1999~XT17 family originate from the breakup of a common parent body.

If the true 1999~XT17 family is composed only by A-types, it could originate from the breakup of an olivine-rich asteroid that was once itself part of a differentiated planetesimal. The latter could have been catastrophically disrupted in another region and a fragment of it could subsequently have been implanted in its current location via a dynamical process \citep{bottke2006,raymond2017}. Corroborating evidence could be the small size (10~km) of asteroid (36256) 1999~XT17, the largest member of the small-sized 1999~XT17 family (some tens of asteroids), coupled with the small diameters of the rest of its members ($\sim$5~km). This was recently shown to be a valid scenario for the parent body of the enstatite meteorites of subtype EL. It has been proposed that this family was delivered to the inner main belt from the terrestrial region during the early evolution phases of our Solar System \citep{Avdellidou2022}. On the other hand, some olivine-rich main belt asteroids could be the result of direct accretion of grains from an oxidized region of the solar nebula, instead of magmatic differentiation, as understood from the study of R-chondrites \citep[]{Schulze1994}. The olivine-rich parent body at the origin of the 1999~XT17 family could have formed through such processes.

%,sunshine2007,sanchez2014,Popescu2018_olivine,demeo2019

%The olivine-rich parent body at the origin of the 1999~XT17 family could also have been formed from nebular processes, \textcolor{purple}{that is to say from the accretion of grains from an oxidized region of the solar nebula}, instead of from magmatic differentiation \citep[]{sunshine2007,sanchez2014,Popescu2018_olivine,demeo2019}.

%The 1999~XT17 family has a V-shape centred around 2.942~au. Using the family's V-shape width $C_0$ = $1.8\times10^{-5}^{-4}$~au from \cite{nesvorny2015}, we notice that every object that we characterised plots inside the family's V-shape. Moreover, every family members having a H$\lesssim$14 and a $S/N>21$ got classified as A-types (see Table \ref{tab:1}), and reside outside the V-shape of Eos. We conclude that, due to their position and their classification, these objects are less likely to be Eos fugitives. 

We go on to explore the hypothesis that the 1999~XT17 family is totally or partially differentiated. As we state above, the second most abundant class in this family according to the literature is the S-complex. S-complex asteroids are typically associated to ordinary chondrite meteorites \citep{vernazza2015,reddy2015,demeo2022}, that are undifferentiated \cite[see][and references therein]{vernazza2015}. Based on the hypothesis where these S-complex objects are correctly classified and belong to the 1999~XT17 family, it is possible that the parent body of the 1999~XT17 family was only partially differentiated. It could have built an olivine mantle without heating enough for the crust to melt, which may have preserved its undifferentiated nature \citep{weiss2013}. %This could explain the presence of S-complex asteroids in the family.
Unfortunately, spectrophotometry is often not precise enough to distinguish between spectroscopic classes. Further spectroscopic observations in VIS and NIR could help in distinguishing between our hypotheses, for example, following the methods of \cite{sunshine2007}.

In any case, spectroscopic studies that attempted to map the olivine-rich asteroids in the main belt concluded that there is no prominent concentration of olivine-rich objects in any asteroid family. They are instead found to be spread throughout the main belt. Here, we identified the first concentration of A-type objects in an asteroid family, using Gaia DR3 data. This identification makes the 1999~XT17 family the first olivine-rich family in the main asteroid belt.

%%%%% add some lines about the meteorites matching in the discussion ?

%-----------------------------------------------------------------
\section{Conclusions}
\label{conclusions}

We report on the discovery of the first concentration of A-type asteroids in an asteroid family in the pristine zone of the outer main belt. The (36256) 1999~XT17 family shows the presence of 16 potential A-type asteroids, among which four have been confirmed by NIR data. The comparison with meteorites of asteroids 36256 and 33763 confirms that these objects should be mostly composed of pure olivine. Depending on the membership in the family of a few S-complex asteroids, we propose two main hypothesis for the origin of the family: (i) the parent body could have been a pure olivine-rich object from the mantle of a partially or totally differentiated planetesimal, which got disrupted possibly at a different location in the Solar System, before getting implanted in the pristine zone; or (ii) the family comes directly from a partially differentiated object that broke in situ. New NIR spectroscopic observations of family members could help disentangle our different scenarios.

%The lack of other spectroscopic class with high abundance among the family's members having a Gaia spectrum, leads to assume that the parent body at the origin of this family was  %New VIS and NIR spectroscopic data could allow to confirm our findings and to further study this family.

%----------------------------------------------------------------- 
\begin{acknowledgements}
MG and MD acknowledge financial support from CNES and the Action Specifique Gaia.
MD, CA, and LG acknowledge financial support from the ANR ORIGINS (ANR-18-CE31-0014).
This work has made use of data from the European Space Agency (ESA) mission
{\it Gaia} (\url{https://www.cosmos.esa.int/gaia}), processed by the {\it Gaia}
Data Processing and Analysis Consortium (DPAC,
\url{https://www.cosmos.esa.int/web/gaia/dpac/consortium}). Funding for the DPAC
has been provided by national institutions, in particular the institutions
participating in the {\it Gaia} Multilateral Agreement. 
The authors made use of the great SpeX instrument at the Infrared Telescope Facility, which is operated by the University of Hawaii under contract 80HQTR19D0030 with the National Aeronautics and Space Administration.
This work is based on data provided by the Minor Planet Physical Properties Catalogue (\url{mp3c.oca.eu}) of the Observatoire de la Côte d'Azur.
\end{acknowledgements}

\bibliographystyle{aa} 
\bibliography{bib.bib}  

\begin{thebibliography}{99}
\expandafter\ifx\csname natexlab\endcsname\relax\def\natexlab#1{#1}\fi

\bibitem[{{Alvarez-Candal} {et~al.}(2006){Alvarez-Candal}, {Duffard},
  {Lazzaro}, \& {Michtchenko}}]{AlvarezCandal2006}
{Alvarez-Candal}, A., {Duffard}, R., {Lazzaro}, D., \& {Michtchenko}, T. 2006,
  \aap, 459, 969

\bibitem[{{Arredondo} {et~al.}(2021){Arredondo}, {Campins}, {Pinilla-Alonso},
  {de Le{\'o}n}, {Lorenzi}, {Morate}, {Rizos}, \& {De Pr{\'a}}}]{Arredondo2021}
{Arredondo}, A., {Campins}, H., {Pinilla-Alonso}, N., {et~al.} 2021, \icarus,
  368, 114619

\bibitem[{{Avdellidou} {et~al.}(2022){Avdellidou}, {Delbo}, {Morbidelli},
  {Walsh}, {Munaibari}, {Bourdelle de Micas}, {Devog{\`e}le}, {Fornasier},
  {Gounelle}, \& {van Belle}}]{Avdellidou2022}
{Avdellidou}, C., {Delbo}, M., {Morbidelli}, A., {et~al.} 2022, \aap, 665, L9

\bibitem[{{Barrat} {et~al.}(2021){Barrat}, {Chaussidon}, {Yamaguchi}, {Beck},
  {Villeneuve}, {Byrne}, {Broadley}, \& {Marty}}]{barrat2021}
{Barrat}, J.-A., {Chaussidon}, M., {Yamaguchi}, A., {et~al.} 2021, Proceedings
  of the National Academy of Science, 118, 2026129118

\bibitem[{{Beck} {et~al.}(2021){Beck}, {Schmitt}, {Potin}, {Pommerol}, \&
  {Brissaud}}]{Beck2021}
{Beck}, P., {Schmitt}, B., {Potin}, S., {Pommerol}, A., \& {Brissaud}, O. 2021,
  \icarus, 354, 114066

\bibitem[{{Binzel} {et~al.}(2004){Binzel}, {Rivkin}, {Stuart}, {Harris}, {Bus},
  \& {Burbine}}]{2004Icar..170..259B}
{Binzel}, R.~P., {Rivkin}, A.~S., {Stuart}, J.~S., {et~al.} 2004, \icarus, 170,
  259

\bibitem[{{Borisov} {et~al.}(2017){Borisov}, {Christou}, {Bagnulo}, {Cellino},
  {Kwiatkowski}, \& {Dell'Oro}}]{Borisov2017}
{Borisov}, G., {Christou}, A., {Bagnulo}, S., {et~al.} 2017, \mnras, 466, 489

\bibitem[{{Bottke} {et~al.}(2006){Bottke}, {Nesvorn{\'y}}, {Grimm},
  {Morbidelli}, \& {O'Brien}}]{bottke2006}
{Bottke}, W.~F., {Nesvorn{\'y}}, D., {Grimm}, R.~E., {Morbidelli}, A., \&
  {O'Brien}, D.~P. 2006, \nat, 439, 821

\bibitem[{{Bro{\v{z}}} \& {Morbidelli}(2019)}]{BrozMorby2019}
{Bro{\v{z}}}, M. \& {Morbidelli}, A. 2019, \icarus, 317, 434

\bibitem[{{Bro{\v{z}}} {et~al.}(2013){Bro{\v{z}}}, {Morbidelli}, {Bottke},
  {Rozehnal}, {Vokrouhlick{\'y}}, \& {Nesvorn{\'y}}}]{Broz2013_pristine}
{Bro{\v{z}}}, M., {Morbidelli}, A., {Bottke}, W.~F., {et~al.} 2013, \aap, 551,
  A117

\bibitem[{{Brunetto} {et~al.}(2006){Brunetto}, {Vernazza}, {Marchi}, {Birlan},
  {Fulchignoni}, {Orofino}, \& {Strazzulla}}]{Brunetto2006}
{Brunetto}, R., {Vernazza}, P., {Marchi}, S., {et~al.} 2006, \icarus, 184, 327

\bibitem[{{Burbine} {et~al.}(1996){Burbine}, {Meibom}, \&
  {Binzel}}]{burbine1996}
{Burbine}, T.~H., {Meibom}, A., \& {Binzel}, R.~P. 1996, \maps, 31, 607

\bibitem[{{Bus} \& {Binzel}(2002)}]{bus2002tax}
{Bus}, S.~J. \& {Binzel}, R.~P. 2002, \icarus, 158, 146

\bibitem[{{Carry}(2012)}]{carry2012}
{Carry}, B. 2012, P\&SS, 73, 98

\bibitem[{{Carvano} {et~al.}(2010){Carvano}, {Hasselmann}, {Lazzaro}, \&
  {Moth{\'e}-Diniz}}]{carvano2010}
{Carvano}, J.~M., {Hasselmann}, P.~H., {Lazzaro}, D., \& {Moth{\'e}-Diniz}, T.
  2010, \aap, 510, A43

\bibitem[{{Carvano} {et~al.}(2001){Carvano}, {Lazzaro}, {Moth{\'e}-Diniz},
  {Angeli}, \& {Florczak}}]{carvano2001}
{Carvano}, J.~M., {Lazzaro}, D., {Moth{\'e}-Diniz}, T., {Angeli}, C.~A., \&
  {Florczak}, M. 2001, \icarus, 149, 173

\bibitem[{{Clark} {et~al.}(2009){Clark}, {Ockert-Bell}, {Cloutis}, {Nesvorny},
  {Moth{\'e}-Diniz}, \& {Bus}}]{clark2009}
{Clark}, B.~E., {Ockert-Bell}, M.~E., {Cloutis}, E.~A., {et~al.} 2009, \icarus,
  202, 119

\bibitem[{{Cloutis} {et~al.}(1990){Cloutis}, {Gaffey}, {Smith}, \&
  {Lambert}}]{cloutis1990}
{Cloutis}, E.~A., {Gaffey}, M.~J., {Smith}, D.~G.~W., \& {Lambert}, R. S.~J.
  1990, \jgr, 95, 8323

\bibitem[{Cruikshank \& Hartmann(1984)}]{cruikshank1984}
Cruikshank, D.~P. \& Hartmann, W.~K. 1984, Science, 223, 281

\bibitem[{{de Le{\'o}n} {et~al.}(2010){de Le{\'o}n}, {Licandro},
  {Serra-Ricart}, {Pinilla-Alonso}, \& {Campins}}]{deleon2010}
{de Le{\'o}n}, J., {Licandro}, J., {Serra-Ricart}, M., {Pinilla-Alonso}, N., \&
  {Campins}, H. 2010, \aap, 517, A23

\bibitem[{{de Le{\'o}n} {et~al.}(2016){de Le{\'o}n}, {Pinilla-Alonso}, {Delbo},
  {Campins}, {Cabrera-Lavers}, {Tanga}, {Cellino}, {Bendjoya}, {Gayon-Markt},
  {Licandro}, {Lorenzi}, {Morate}, {Walsh}, {DeMeo}, {Landsman}, \&
  {Al{\'\i}-Lagoa}}]{2016Icar..266...57D}
{de Le{\'o}n}, J., {Pinilla-Alonso}, N., {Delbo}, M., {et~al.} 2016, \icarus,
  266, 57

\bibitem[{{De Pr{\'a}} {et~al.}(2020{\natexlab{a}}){De Pr{\'a}}, {Licandro},
  {Pinilla-Alonso}, {Lorenzi}, {Rond{\'o}n}, {Carvano}, {Morate}, \& {De
  Le{\'o}n}}]{depra2020}
{De Pr{\'a}}, M.~N., {Licandro}, J., {Pinilla-Alonso}, N., {et~al.}
  2020{\natexlab{a}}, \icarus, 338, 113473

\bibitem[{{De Pr{\'a}} {et~al.}(2020{\natexlab{b}}){De Pr{\'a}},
  {Pinilla-Alonso}, {Carvano}, {Licandro}, {Morate}, {Lorenzi}, {de Le{\'o}n},
  {Campins}, \& {Moth{\'e}-Diniz}}]{depra2020b}
{De Pr{\'a}}, M.~N., {Pinilla-Alonso}, N., {Carvano}, J., {et~al.}
  2020{\natexlab{b}}, \aap, 643, A102

\bibitem[{{De Pr{\'a}} {et~al.}(2018){De Pr{\'a}}, {Pinilla-Alonso}, {Carvano},
  {Licandro}, {Campins}, {Moth{\'e}-Diniz}, {De Le{\'o}n}, \&
  {Al{\'\i}-Lagoa}}]{depra2018}
{De Pr{\'a}}, M.~N., {Pinilla-Alonso}, N., {Carvano}, J.~M., {et~al.} 2018,
  \icarus, 311, 35

\bibitem[{{de Sanctis} {et~al.}(2011){de Sanctis}, {Migliorini}, {Luzia
  Jasmin}, {Lazzaro}, {Filacchione}, {Marchi}, {Ammannito}, \&
  {Capria}}]{DeSanctis2011}
{de Sanctis}, M.~C., {Migliorini}, A., {Luzia Jasmin}, F., {et~al.} 2011, \aap,
  533, A77

\bibitem[{{Delbo'} {et~al.}(2012){Delbo'}, {Gayon-Markt}, {Busso}, {Brown},
  {Galluccio}, {Ordenovic}, {Bendjoya}, \& {Tanga}}]{delbo2012}
{Delbo'}, M., {Gayon-Markt}, J., {Busso}, G., {et~al.} 2012, \planss, 73, 86

\bibitem[{{DeMeo} {et~al.}(2014){DeMeo}, {Binzel}, {Carry}, {Polishook}, \&
  {Moskovitz}}]{demeo2014}
{DeMeo}, F.~E., {Binzel}, R.~P., {Carry}, B., {Polishook}, D., \& {Moskovitz},
  N.~A. 2014, \icarus, 229, 392

\bibitem[{{DeMeo} {et~al.}(2009){DeMeo}, {Binzel}, {Slivan}, \&
  {Bus}}]{demeo2009}
{DeMeo}, F.~E., {Binzel}, R.~P., {Slivan}, S.~M., \& {Bus}, S.~J. 2009, Icarus,
  202, 160

\bibitem[{{DeMeo} {et~al.}(2022){DeMeo}, {Burt}, {Marsset}, {Polishook},
  {Burbine}, {Carry}, {Binzel}, {Vernazza}, {Reddy}, {Tang}, {Thomas},
  {Rivkin}, {Moskovitz}, {Slivan}, \& {Bus}}]{demeo2022}
{DeMeo}, F.~E., {Burt}, B.~J., {Marsset}, M., {et~al.} 2022, \icarus, 380,
  114971

\bibitem[{{DeMeo} \& {Carry}(2013)}]{demeo2013}
{DeMeo}, F.~E. \& {Carry}, B. 2013, \icarus, 226, 723

\bibitem[{{DeMeo} {et~al.}(2019){DeMeo}, {Polishook}, {Carry}, {Burt}, {Hsieh},
  {Binzel}, {Moskovitz}, \& {Burbine}}]{demeo2019}
{DeMeo}, F.~E., {Polishook}, D., {Carry}, B., {et~al.} 2019, \icarus, 322, 13

\bibitem[{{Devog{\`e}le} {et~al.}(2019){Devog{\`e}le}, {Moskovitz}, {Thirouin},
  {Gustaffson}, {Magnuson}, {Thomas}, {Willman}, {Christensen}, {Person},
  {Binzel}, {Polishook}, {DeMeo}, {Hinkle}, {Trilling}, {Mommert}, {Burt}, \&
  {Skiff}}]{devogele2019}
{Devog{\`e}le}, M., {Moskovitz}, N., {Thirouin}, A., {et~al.} 2019, \aj, 158,
  196

\bibitem[{{Devog{\`e}le} {et~al.}(2018){Devog{\`e}le}, {Tanga}, {Cellino},
  {Bendjoya}, {Rivet}, {Surdej}, {Vernet}, {Sunshine}, {Bus}, {Abe}, {Bagnulo},
  {Borisov}, {Campins}, {Carry}, {Licandro}, {McLean}, \&
  {Pinilla-Alonso}}]{devogele2018}
{Devog{\`e}le}, M., {Tanga}, P., {Cellino}, A., {et~al.} 2018, \icarus, 304, 31

\bibitem[{{Duffard} {et~al.}(2004){Duffard}, {Lazzaro}, {Licandro}, {De
  Sanctis}, {Capria}, \& {Carvano}}]{2004Icar..171..120D}
{Duffard}, R., {Lazzaro}, D., {Licandro}, J., {et~al.} 2004, \icarus, 171, 120

\bibitem[{{Durech} {et~al.}(2010){Durech}, {Sidorin}, \&
  {Kaasalainen}}]{durech2010}
{Durech}, J., {Sidorin}, V., \& {Kaasalainen}, M. 2010, \aap, 513, A46

\bibitem[{{Elkins-Tanton} \& {Weiss}(2017)}]{elkins-tanton2017}
{Elkins-Tanton}, L.~T. \& {Weiss}, B.~P. 2017, {Planetesimals: Early
  Differentiation and Consequences for Planets}

\bibitem[{{Fienga} {et~al.}(2020){Fienga}, {Avdellidou}, \&
  {Hanu{\v{s}}}}]{fienga2020}
{Fienga}, A., {Avdellidou}, C., \& {Hanu{\v{s}}}, J. 2020, \mnras, 492, 589

\bibitem[{{Fornasier} {et~al.}(2010){Fornasier}, {Clark}, {Dotto},
  {Migliorini}, {Ockert-Bell}, \& {Barucci}}]{fornasier2010}
{Fornasier}, S., {Clark}, B.~E., {Dotto}, E., {et~al.} 2010, \icarus, 210, 655

\bibitem[{{Gaia Collaboration} {et~al.}(2022){Gaia Collaboration}, {Galluccio},
  {Delbo}, {De Angeli}, {Pauwels}, {Tanga}, {Mignard}, {Cellino}, {Brown},
  {Muinonen}, {Penttila}, {Jordan}, {Vallenari}, {Prusti}, {de Bruijne},
  {Arenou}, {Babusiaux}, {Biermann}, {Creevey}, {Ducourant}, {Evans}, {Eyer},
  {Guerra}, {Hutton}, {Jordi}, {Klioner}, {Lammers}, {Lindegren}, {Luri},
  {Panem}, {Pourbaix}, {Randich}, {Sartoretti}, {Soubiran}, {Walton},
  {Bailer-Jones}, {Bastian}, {Drimmel}, {Jansen}, {Katz}, {Lattanzi}, {van
  Leeuwen}, {Bakker}, {Cacciari}, {Castaneda}, {Fabricius}, {Fouesneau},
  {Fr{\'e}mat}, {Guerrier}, {Heiter}, {Masana}, {Messineo}, {Mowlavi},
  {Nicolas}, {Nienartowicz}, {Pailler}, {Panuzzo}, {Riclet}, {Roux},
  {Seabroke}, {Sordo}, {Th{\'e}venin}, {Gracia-Abril}, {Portell}, {Teyssier},
  {Altmann}, {Andrae}, {Audard}, {Bellas-Velidis}, {Benson}, {Berthier},
  {Blomme}, {Burgess}, {Busonero}, {Busso}, {C{\'a}novas}, {Carry}, {Cheek},
  {Clementini}, {Damerdji}, {Davidson}, {de Teodoro}, {Nunez Campos},
  {Delchambre}, {Dell Oro}, {Esquej}, {Fern{\'a}ndez-Hern{\'a}ndez}, {Fraile},
  {Garabato}, {Garc{\'\i}a-Lario}, {Gosset}, {Haigron}, {Halbwachs}, {Hambly},
  {Harrison}, {Hern{\'a}ndez}, {Hestroffer}, {Hodgkin}, {Holl}, {Janssen},
  {Jevardat de Fombelle}, {Krone-Martins}, {Lanzafame}, {L{\"o}ffler},
  {Marchal}, {Marrese}, {Moitinho}, {Osborne}, {Pancino}, {Recio-Blanco},
  {Reyl{\'e}}, {Riello}, {Rimoldini}, {Roegiers}, {Rybizki}, {Sarro}, {Siopis},
  {Smith}, {Sozzetti}, {Utrilla}, {van Leeuwen}, {Abbas}, {{\'A}brah{\'a}m},
  {Abreu Aramburu}, {Aerts}, {Aguado}, {Ajaj}, {Aldea-Montero}, {Altavilla},
  {{\'A}lvarez}, {Alves}, {Anderson}, {Anglada Varela}, {Antoja}, {Baines},
  {Baker}, {Balaguer-N{\'u}nez}, {Balbinot}, {Balog}, {Barache}, {Barbato},
  {Barros}, {Barstow}, {Bartolom{\'e}}, {Bassilana}, {Bauchet}, {Becciani},
  {Bellazzini}, {Berihuete}, {Bernet}, {Bertone}, {Bianchi}, {Binnenfeld},
  {Blanco-Cuaresma}, {Boch}, {Bombrun}, {Bossini}, {Bouquillon}, {Bragaglia},
  {Bramante}, {Breedt}, {Bressan}, {Brouillet}, {Brugaletta}, {Bucciarelli},
  {Burlacu}, {Butkevich}, {Buzzi}, {Caffau}, {Cancelliere}, {Cantat-Gaudin},
  {Carballo}, {Carlucci}, {Carnerero}, {Carrasco}, {Casamiquela}, {Castellani},
  {Castro-Ginard}, {Chaoul}, {Charlot}, {Chemin}, {Chiaramida}, {Chiavassa},
  {Chornay}, {Comoretto}, {Contursi}, {Cooper}, {Cornez}, {Cowell}, {Crifo},
  {Cropper}, {Crosta}, {Crowley}, {Dafonte}, {Dapergolas}, {David}, {de
  Laverny}, {De Luise}, {De March}, {De Ridder}, {de Souza}, {de Torres}, {del
  Peloso}, {del Pozo}, {Delgado}, {Delisle}, {Demouchy}, {Dharmawardena},
  {Diakite}, {Diener}, {Distefano}, {Dolding}, {Enke}, {Fabre}, {Fabrizio},
  {Faigler}, {Fedorets}, {Fernique}, {Figueras}, {Fournier}, {Fouron},
  {Fragkoudi}, {Gai}, {Garcia-Gutierrez}, {Garcia-Reinaldos},
  {Garc{\'\i}a-Torres}, {Garofalo}, {Gavel}, {Gavras}, {Gerlach}, {Geyer},
  {Giacobbe}, {Gilmore}, {Girona}, {Giuffrida}, {Gomel}, {Gomez},
  {Gonz{\'a}lez-N{\'u}nez}, {Gonz{\'a}lez-Santamar{\'\i}a},
  {Gonz{\'a}lez-Vidal}, {Granvik}, {Guillout}, {Guiraud},
  {Guti{\'e}rrez-S{\'a}nchez}, {Guy}, {Hatzidimitriou}, {Hauser}, {Haywood},
  {Helmer}, {Helmi}, {Sarmiento}, {Hidalgo}, {Hadczuk}, {Hobbs}, {Holland},
  {Huckle}, {Jardine}, {Jasniewicz}, {Jean-Antoine Piccolo},
  {Jim{\'e}nez-Arranz}, {Juaristi Campillo}, {Julbe}, {Karbevska}, {Kervella},
  {Khanna}, {Kordopatis}, {Korn}, {Kosp{\'a}l}, {Kostrzewa-Rutkowska},
  {Kruszynska}, {Kun}, {Laizeau}, {Lambert}, {Lanza}, {Lasne}, {Le Campion},
  {Lebreton}, {Lebzelter}, {Leccia}, {Leclerc}, {Lecoeur-Taibi}, {Liao},
  {Licata}, {Lindstrom}, {Lister}, {Livanou}, {Lobel}, {Lorca}, {Loup},
  {Madrero Pardo}, {Magdaleno Romeo}, {Managau}, {Mann}, {Manteiga},
  {Marchant}, {Marconi}, {Marcos}, {Marcos Santos}, {Mar{\'\i}n Pina},
  {Marinoni}, {Marocco}, {Marshall}, {Polo}, {Mart{\'\i}n-Fleitas}, {Marton},
  {Mary}, {Masip}, {Massari}, {Mastrobuono-Battisti}, {Mazeh}, {McMillan},
  {Messina}, {Michalik}, {Millar}, {Mints}, {Molina}, {Molinaro}, {Moln{\'a}r},
  {Monari}, {Mongui{\'o}}, {Montegriffo}, {Montero}, {Mor}, {Mora},
  {Morbidelli}, {Morel}, {Morris}, {Muraveva}, {Murphy}, {Musella}, {Nagy},
  {Noval}, {Ocana}, {Ogden}, {Ordenovic}, {Osinde}, {Pagani}, {Pagano},
  {Palaversa}, {Palicio}, {Pallas-Quintela}, {Panahi}, {Payne-Wardenaar},
  {Penalosa Esteller}, {Petit}, {Pichon}, {Piersimoni}, {Pineau}, {Plachy},
  {Plum}, {Poggio}, {Prsa}, {Pulone}, {Racero}, {Ragaini}, {Rainer}, {Raiteri},
  {Ramos}, {Ramos-Lerate}, {Re Fiorentin}, {Regibo}, {Richards}, {Rios Diaz},
  {Ripepi}, {Riva}, {Rix}, {Rixon}, {Robichon}, {Robin}, {Robin}, {Roelens},
  {Rogues}, {Rohrbasser}, {Romero-G{\'o}mez}, {Rowell}, {Royer}, {Ruz Mieres},
  {Rybicki}, {Sadowski}, {S{\'a}ez N{\'u}nez}, {Sagrist{\`a} Sell{\'e}s},
  {Sahlmann}, {Salguero}, {Samaras}, {Sanchez Gimenez}, {Sanna}, {Santovena},
  {Sarasso}, {Schultheis}, {Sciacca}, {Segol}, {Segovia}, {S{\'e}gransan},
  {Semeux}, {Shahaf}, {Siddiqui}, {Siebert}, {Siltala}, {Silvelo}, {Slezak},
  {Slezak}, {Smart}, {Snaith}, {Solano}, {Solitro}, {Souami}, {Souchay},
  {Spagna}, {Spina}, {Spoto}, {Steele}, {Steidelm{\"u}ller}, {Stephenson},
  {S{\"u}veges}, {Surdej}, {Szabados}, {Szegedi-Elek}, {Taris}, {Taylor},
  {Teixeira}, {Tolomei}, {Tonello}, {Torra}, {Torra}, {Torralba Elipe},
  {Trabucchi}, {Tsounis}, {Turon}, {Ulla}, {Unger}, {Vaillant}, {van Dillen},
  {van Reeven}, {Vanel}, {Vecchiato}, {Viala}, {Vicente}, {Voutsinas},
  {Weiler}, {Wevers}, {Wyrzykowski}, {Yoldas}, {Yvard}, {Zhao}, {Zorec},
  {Zucker}, \& {Zwitter}}]{galluccio2022}
{Gaia Collaboration}, {Galluccio}, L., {Delbo}, M., {et~al.} 2022, arXiv
  e-prints, arXiv:2206.12174

\bibitem[{{Galinier} {et~al.}(2023){Galinier}, {Delbo}, {Avdellidou},
  {Galluccio}, \& {Marrocchi}}]{galinier2023}
{Galinier}, M., {Delbo}, M., {Avdellidou}, C., {Galluccio}, L., \& {Marrocchi},
  Y. 2023, \aap, 671, A40

\bibitem[{{Gartrelle} {et~al.}(2021){Gartrelle}, {Hardersen}, {Izawa}, \&
  {Nowinski}}]{Gartrelle2021}
{Gartrelle}, G.~M., {Hardersen}, P.~S., {Izawa}, M. R.~M., \& {Nowinski}, M.~C.
  2021, \icarus, 363, 114295

\bibitem[{{Gietzen} {et~al.}(2012){Gietzen}, {Lacy}, {Ostrowski}, \&
  {Sears}}]{Gietzen2012}
{Gietzen}, K.~M., {Lacy}, C. H.~S., {Ostrowski}, D.~R., \& {Sears}, D. W.~G.
  2012, \maps, 47, 1789

\bibitem[{{Greenwood} {et~al.}(2006){Greenwood}, {Franchi}, {Jambon}, {Barrat},
  \& {Burbine}}]{greenwood2006}
{Greenwood}, R.~C., {Franchi}, I.~A., {Jambon}, A., {Barrat}, J.~A., \&
  {Burbine}, T.~H. 2006, Science, 313, 1763

\bibitem[{{Gulbis} {et~al.}(2011){Gulbis}, {Bus}, {Elliot}, {Rayner},
  {Stahlberger}, {Rojas}, {Adams}, {Person}, {Chung}, {Tokunaga}, \&
  {Zuluaga}}]{gulbis2011}
{Gulbis}, A.~A.~S., {Bus}, S.~J., {Elliot}, J.~L., {et~al.} 2011, \pasp, 123,
  461

\bibitem[{{Harris} \& {Drube}(2014)}]{harris&drube2014}
{Harris}, A.~W. \& {Drube}, L. 2014, \apjl, 785, L4

\bibitem[{{Hasegawa} {et~al.}(2021){Hasegawa}, {Marsset}, {DeMeo}, {Bus},
  {Geem}, {Ishiguro}, {Im}, {Kuroda}, \& {Vernazza}}]{Hasegawa2021}
{Hasegawa}, S., {Marsset}, M., {DeMeo}, F.~E., {et~al.} 2021, \apjl, 916, L6

\bibitem[{{Hewins}(1983)}]{hewins1983}
{Hewins}, R.~H. 1983, Lunar and Planetary Science Conference Proceedings, 88,
  B257

\bibitem[{{Kruijer} {et~al.}(2014){Kruijer}, {Touboul}, {Fischer-G{\"o}dde},
  {Bermingham}, {Walker}, \& {Kleine}}]{kruijer2014}
{Kruijer}, T.~S., {Touboul}, M., {Fischer-G{\"o}dde}, M., {et~al.} 2014,
  Science, 344, 1150

\bibitem[{{Lazzarin} {et~al.}(2004){Lazzarin}, {Marchi}, {Barucci}, {Di
  Martino}, \& {Barbieri}}]{Lazzarin2004}
{Lazzarin}, M., {Marchi}, S., {Barucci}, M.~A., {Di Martino}, M., \&
  {Barbieri}, C. 2004, \icarus, 169, 373

\bibitem[{{Lazzarin} {et~al.}(2005){Lazzarin}, {Marchi}, {Magrin}, \&
  {Licandro}}]{Lazzarin2005}
{Lazzarin}, M., {Marchi}, S., {Magrin}, S., \& {Licandro}, J. 2005, \mnras,
  359, 1575

\bibitem[{{Leith} {et~al.}(2017){Leith}, {Moskovitz}, {Mayne}, {DeMeo},
  {Takir}, {Burt}, {Binzel}, \& {Pefkou}}]{leith2017}
{Leith}, T.~B., {Moskovitz}, N.~A., {Mayne}, R.~G., {et~al.} 2017, \icarus,
  295, 61

\bibitem[{{Licandro} {et~al.}(2008){Licandro}, {Alvarez-Candal}, {de Le{\'o}n},
  {Pinilla-Alonso}, {Lazzaro}, \& {Campins}}]{2008A&A...487.1195L}
{Licandro}, J., {Alvarez-Candal}, A., {de Le{\'o}n}, J., {et~al.} 2008, \aap,
  487, 1195

\bibitem[{{Lucas} {et~al.}(2019){Lucas}, {Emery}, {MacLennan},
  {Pinilla-Alonso}, {Cartwright}, {Lindsay}, {Reddy}, {Sanchez}, {Thomas}, \&
  {Lorenzi}}]{lucas2019}
{Lucas}, M.~P., {Emery}, J.~P., {MacLennan}, E.~M., {et~al.} 2019, \icarus,
  322, 227

\bibitem[{{Lucas} {et~al.}(2017){Lucas}, {Emery}, {Pinilla-Alonso}, {Lindsay},
  \& {Lorenzi}}]{lucas2017}
{Lucas}, M.~P., {Emery}, J.~P., {Pinilla-Alonso}, N., {Lindsay}, S.~S., \&
  {Lorenzi}, V. 2017, \icarus, 291, 268

\bibitem[{{Mahlke} {et~al.}(2022){Mahlke}, {Carry}, \& {Mattei}}]{Mahlke2022}
{Mahlke}, M., {Carry}, B., \& {Mattei}, P.~A. 2022, \aap, 665, A26

\bibitem[{{Mahlke} {et~al.}(2023){Mahlke}, {Eschrig}, {Carry}, {Bonal}, \&
  {Beck}}]{Mahlke2023}
{Mahlke}, M., {Eschrig}, J., {Carry}, B., {Bonal}, L., \& {Beck}, P. 2023,
  \aap, 676, A94

\bibitem[{{Marchi} {et~al.}(2005){Marchi}, {Lazzarin}, {Paolicchi}, \&
  {Magrin}}]{2005Icar..175..170M}
{Marchi}, S., {Lazzarin}, M., {Paolicchi}, P., \& {Magrin}, S. 2005, \icarus,
  175, 170

\bibitem[{{Marchi} {et~al.}(2012){Marchi}, {McSween}, {O'Brien}, {Schenk}, {De
  Sanctis}, {Gaskell}, {Jaumann}, {Mottola}, {Preusker}, {Raymond}, {Roatsch},
  \& {Russell}}]{marchi2012}
{Marchi}, S., {McSween}, H.~Y., {O'Brien}, D.~P., {et~al.} 2012, Science, 336,
  690

\bibitem[{{McCord} {et~al.}(1970){McCord}, {Adams}, \& {Johnson}}]{mccord1970}
{McCord}, T.~B., {Adams}, J.~B., \& {Johnson}, T.~V. 1970, Science, 168, 1445

\bibitem[{{Migliorini} {et~al.}(2017){Migliorini}, {De Sanctis}, {Lazzaro}, \&
  {Ammannito}}]{migliorini2017}
{Migliorini}, A., {De Sanctis}, M.~C., {Lazzaro}, D., \& {Ammannito}, E. 2017,
  \mnras, 464, 1718

\bibitem[{{Migliorini} {et~al.}(2018){Migliorini}, {De Sanctis}, {Lazzaro}, \&
  {Ammannito}}]{migliorini2018}
{Migliorini}, A., {De Sanctis}, M.~C., {Lazzaro}, D., \& {Ammannito}, E. 2018,
  \mnras, 475, 353

\bibitem[{{Migliorini} {et~al.}(2021){Migliorini}, {De Sanctis}, {Michtchenko},
  {Lazzaro}, {Barbieri}, {Mesa}, {Lazzarin}, \& {La Forgia}}]{Migliorini2021}
{Migliorini}, A., {De Sanctis}, M.~C., {Michtchenko}, T.~A., {et~al.} 2021,
  \mnras, 504, 2019

\bibitem[{{Morate} {et~al.}(2018){Morate}, {de Le{\'o}n}, {De Pr{\'a}},
  {Licandro}, {Cabrera-Lavers}, {Campins}, \& {Pinilla-Alonso}}]{morate2018}
{Morate}, D., {de Le{\'o}n}, J., {De Pr{\'a}}, M., {et~al.} 2018, \aap, 610,
  A25

\bibitem[{{Morate} {et~al.}(2016){Morate}, {de Le{\'o}n}, {De Pr{\'a}},
  {Licandro}, {Cabrera-Lavers}, {Campins}, {Pinilla-Alonso}, \&
  {Al{\'\i}-Lagoa}}]{2016A&A...586A.129M}
{Morate}, D., {de Le{\'o}n}, J., {De Pr{\'a}}, M., {et~al.} 2016, \aap, 586,
  A129

\bibitem[{{Morate} {et~al.}(2019){Morate}, {de Le{\'o}n}, {De Pr{\'a}},
  {Licandro}, {Pinilla-Alonso}, {Campins}, {Arredondo}, {Carvano}, {Lazzaro},
  \& {Cabrera-Lavers}}]{morate2019}
{Morate}, D., {de Le{\'o}n}, J., {De Pr{\'a}}, M., {et~al.} 2019, \aap, 630,
  A141

\bibitem[{{Moth{\'e}-Diniz} {et~al.}(2008){Moth{\'e}-Diniz}, {Carvano}, {Bus},
  {Duffard}, \& {Burbine}}]{2008Icar..195..277M}
{Moth{\'e}-Diniz}, T., {Carvano}, J.~M., {Bus}, S.~J., {Duffard}, R., \&
  {Burbine}, T.~H. 2008, \icarus, 195, 277

\bibitem[{{Moth{\'e}-Diniz} \& {Nesvorn{\'y}}(2008)}]{2008A&A...486L...9M}
{Moth{\'e}-Diniz}, T. \& {Nesvorn{\'y}}, D. 2008, \aap, 486, L9

\bibitem[{{Moth{\'e}-Diniz} {et~al.}(2005){Moth{\'e}-Diniz}, {Roig}, \&
  {Carvano}}]{2005Icar..174...54M}
{Moth{\'e}-Diniz}, T., {Roig}, F., \& {Carvano}, J.~M. 2005, \icarus, 174, 54

\bibitem[{{Nesvorny} {et~al.}(2015){Nesvorny}, {Broz}, \&
  {Carruba}}]{nesvorny2015}
{Nesvorny}, D., {Broz}, M., \& {Carruba}, V. 2015, in Asteroids IV (P. Michel,
  et al. eds), 297

\bibitem[{{Oszkiewicz} {et~al.}(2023){Oszkiewicz}, {Klimczak}, {Carry},
  {Penttil{\"a}}, {Popescu}, {Kr{\"u}ger}, \& {Aron Keniger}}]{Oszkiewicz2023}
{Oszkiewicz}, D., {Klimczak}, H., {Carry}, B., {et~al.} 2023, \mnras, 519, 2917

\bibitem[{{Oszkiewicz} {et~al.}(2014){Oszkiewicz}, {Kwiatkowski}, {Tomov},
  {Birlan}, {Geier}, {Penttila}, \& {Polinska}}]{Oszkiewicz2014}
{Oszkiewicz}, D.~A., {Kwiatkowski}, T., {Tomov}, T., {et~al.} 2014, \aap, 572,
  A29

\bibitem[{{Perna} {et~al.}(2018){Perna}, {Barucci}, {Fulchignoni}, {Popescu},
  {Belskaya}, {Fornasier}, {Doressoundiram}, {Lantz}, \& {Merlin}}]{perna2018}
{Perna}, D., {Barucci}, M.~A., {Fulchignoni}, M., {et~al.} 2018, \planss, 157,
  82

\bibitem[{{Popescu} {et~al.}(2012){Popescu}, {Birlan}, \&
  {Nedelcu}}]{popescu2012}
{Popescu}, M., {Birlan}, M., \& {Nedelcu}, D.~A. 2012, \aap, 544, A130

\bibitem[{{Popescu} {et~al.}(2014){Popescu}, {Birlan}, {Nedelcu}, {Vaubaillon},
  \& {Cristescu}}]{2014A&A...572A.106P}
{Popescu}, M., {Birlan}, M., {Nedelcu}, D.~A., {Vaubaillon}, J., \&
  {Cristescu}, C.~P. 2014, \aap, 572, A106

\bibitem[{{Popescu} {et~al.}(2018{\natexlab{a}}){Popescu}, {Licandro},
  {Carvano}, {Stoicescu}, {de Le{\'o}n}, {Morate}, {Boac{\u{a}}}, \&
  {Cristescu}}]{Popescu_2018_classification}
{Popescu}, M., {Licandro}, J., {Carvano}, J.~M., {et~al.} 2018{\natexlab{a}},
  \aap, 617, A12

\bibitem[{{Popescu} {et~al.}(2018{\natexlab{b}}){Popescu}, {Perna}, {Barucci},
  {Fornasier}, {Doressoundiram}, {Lantz}, {Merlin}, {Belskaya}, \&
  {Fulchignoni}}]{Popescu2018_olivine}
{Popescu}, M., {Perna}, D., {Barucci}, M.~A., {et~al.} 2018{\natexlab{b}},
  \mnras, 477, 2786

\bibitem[{{Popescu} {et~al.}(2019){Popescu}, {Vaduvescu}, {de Le{\'o}n},
  {Gherase}, {Licandro}, {Boac{\u{a}}}, {{\c{S}}onka}, {Ashley},
  {Mo{\v{c}}nik}, {Morate}, {Predatu}, {De Pr{\'a}}, {Fari{\~n}a}, {Stoev},
  {D{\'\i}az Alfaro}, {Ordonez-Etxeberria}, {L{\'o}pez-Mart{\'\i}nez}, \&
  {Errmann}}]{popescu2019}
{Popescu}, M., {Vaduvescu}, O., {de Le{\'o}n}, J., {et~al.} 2019, \aap, 627,
  A124

\bibitem[{{Raymond} \& {Izidoro}(2017)}]{raymond2017}
{Raymond}, S.~N. \& {Izidoro}, A. 2017, Science Advances, 3, e1701138

\bibitem[{{Rayner} {et~al.}(2003){Rayner}, {Toomey}, {Onaka}, {Denault},
  {Stahlberger}, {Vacca}, {Cushing}, \& {Wang}}]{rayner2003}
{Rayner}, J.~T., {Toomey}, D.~W., {Onaka}, P.~M., {et~al.} 2003, \pasp, 115,
  362

\bibitem[{{Reddy} {et~al.}(2015){Reddy}, {Dunn}, {Thomas}, {Moskovitz}, \&
  {Burbine}}]{reddy2015}
{Reddy}, V., {Dunn}, T.~L., {Thomas}, C.~A., {Moskovitz}, N.~A., \& {Burbine},
  T.~H. 2015, in Asteroids IV (P. Michel, et al. eds), 43

\bibitem[{{Reddy} {et~al.}(2009){Reddy}, {Emery}, {Gaffey}, {Bottke}, {Cramer},
  \& {Kelley}}]{reddy2009}
{Reddy}, V., {Emery}, J.~P., {Gaffey}, M.~J., {et~al.} 2009, \maps, 44, 1917

\bibitem[{{Ribeiro} {et~al.}(2014){Ribeiro}, {Roig}, {Ca{\~n}ada-Assandri},
  {Carvano}, {Jasmin}, {Alvarez-Candal}, \& {Gil-Hutton}}]{Riberio2014}
{Ribeiro}, A.~O., {Roig}, F., {Ca{\~n}ada-Assandri}, M., {et~al.} 2014,
  \planss, 92, 57

\bibitem[{{Roig} {et~al.}(2008){Roig}, {Nesvorn{\'y}}, {Gil-Hutton}, \&
  {Lazzaro}}]{2008Icar..194..125R}
{Roig}, F., {Nesvorn{\'y}}, D., {Gil-Hutton}, R., \& {Lazzaro}, D. 2008,
  \icarus, 194, 125

\bibitem[{{Russell} {et~al.}(2012){Russell}, {Raymond}, {Coradini}, {McSween},
  {Zuber}, {Nathues}, {De Sanctis}, {Jaumann}, {Konopliv}, {Preusker}, {Asmar},
  {Park}, {Gaskell}, {Keller}, {Mottola}, {Roatsch}, {Scully}, {Smith},
  {Tricarico}, {Toplis}, {Christensen}, {Feldman}, {Lawrence}, {McCoy},
  {Prettyman}, {Reedy}, {Sykes}, \& {Titus}}]{russell2012}
{Russell}, C.~T., {Raymond}, C.~A., {Coradini}, A., {et~al.} 2012, Science,
  336, 684

\bibitem[{{Sanchez} {et~al.}(2013){Sanchez}, {Michelsen}, {Reddy}, \&
  {Nathues}}]{Sanchez2013}
{Sanchez}, J.~A., {Michelsen}, R., {Reddy}, V., \& {Nathues}, A. 2013, \icarus,
  225, 131

\bibitem[{{Sanchez} {et~al.}(2014){Sanchez}, {Reddy}, {Kelley}, {Cloutis},
  {Bottke}, {Nesvorn{\'y}}, {Lucas}, {Hardersen}, {Gaffey}, {Abell}, \&
  {Corre}}]{sanchez2014}
{Sanchez}, J.~A., {Reddy}, V., {Kelley}, M.~S., {et~al.} 2014, Icarus, 228, 288

\bibitem[{{Schulze} {et~al.}(1994){Schulze}, {Bischoff}, {Palme}, {Spettel},
  {Dreibus}, \& {Otto}}]{Schulze1994}
{Schulze}, H., {Bischoff}, A., {Palme}, H., {et~al.} 1994, Meteoritics, 29, 275

\bibitem[{{Sergeyev} \& {Carry}(2021)}]{SergeyevCarry2021}
{Sergeyev}, A.~V. \& {Carry}, B. 2021, \aap, 652, A59

\bibitem[{{Sergeyev} {et~al.}(2022){Sergeyev}, {Carry}, {Onken}, {Devillepoix},
  {Wolf}, \& {Chang}}]{Sergeyev2022}
{Sergeyev}, A.~V., {Carry}, B., {Onken}, C.~A., {et~al.} 2022, \aap, 658, A109

\bibitem[{{Sunshine} {et~al.}(2007){Sunshine}, {Bus}, {Corrigan}, {McCoy}, \&
  {Burbine}}]{sunshine2007}
{Sunshine}, J.~M., {Bus}, S.~J., {Corrigan}, C.~M., {McCoy}, T.~J., \&
  {Burbine}, T.~H. 2007, \maps, 42, 155

\bibitem[{{Tatsumi} {et~al.}(2022){Tatsumi}, {Tinaut-Ruano}, {de Le{\'o}n},
  {Popescu}, \& {Licandro}}]{tatsumi2022}
{Tatsumi}, E., {Tinaut-Ruano}, F., {de Le{\'o}n}, J., {Popescu}, M., \&
  {Licandro}, J. 2022, \aap, 664, A107

\bibitem[{{Thomas} {et~al.}(2014){Thomas}, {Emery}, {Trilling}, {Delb{\'o}},
  {Hora}, \& {Mueller}}]{2014Icar..228..217T}
{Thomas}, C.~A., {Emery}, J.~P., {Trilling}, D.~E., {et~al.} 2014, \icarus,
  228, 217

\bibitem[{{Tsirvoulis} {et~al.}(2018){Tsirvoulis}, {Morbidelli}, {Delbo}, \&
  {Tsiganis}}]{tsirvoulis2018}
{Tsirvoulis}, G., {Morbidelli}, A., {Delbo}, M., \& {Tsiganis}, K. 2018,
  \icarus, 304, 14

\bibitem[{{{\v D}urech} {et~al.}(2011){{\v D}urech}, {Kaasalainen}, {Herald},
  {Dunham}, {Timerson}, {Hanu{\v s}}, {Frappa}, {Talbot}, {Hayamizu}, {Warner},
  {Pilcher}, \& {Gal{\'a}d}}]{durech2011}
{{\v D}urech}, J., {Kaasalainen}, M., {Herald}, D., {et~al.} 2011, Icarus, 214,
  652

\bibitem[{{Vernazza} {et~al.}(2021){Vernazza}, {Ferrais}, {Jorda},
  {Hanu{\v{s}}}, {Carry}, {Marsset}, {Bro{\v{z}}}, {Fetick}, {Viikinkoski},
  {Marchis}, {Vachier}, {Drouard}, {Fusco}, {Birlan}, {Podlewska-Gaca},
  {Rambaux}, {Neveu}, {Bartczak}, {Dudzi{\'n}ski}, {Jehin}, {Beck}, {Berthier},
  {Castillo-Rogez}, {Cipriani}, {Colas}, {Dumas}, {{\v{D}}urech}, {Grice},
  {Kaasalainen}, {Kryszczynska}, {Lamy}, {Le Coroller}, {Marciniak},
  {Michalowski}, {Michel}, {Santana-Ros}, {Tanga}, {Vigan}, {Witasse}, {Yang},
  {Antonini}, {Audejean}, {Aurard}, {Behrend}, {Benkhaldoun}, {Bosch},
  {Chapman}, {Dalmon}, {Fauvaud}, {Hamanowa}, {Hamanowa}, {His}, {Jones},
  {Kim}, {Kim}, {Krajewski}, {Labrevoir}, {Leroy}, {Livet}, {Molina},
  {Montaigut}, {Oey}, {Payre}, {Reddy}, {Sabin}, {Sanchez}, \&
  {Socha}}]{vernazza2021}
{Vernazza}, P., {Ferrais}, M., {Jorda}, L., {et~al.} 2021, \aap, 654, A56

\bibitem[{{Vernazza} {et~al.}(2015){Vernazza}, {Zanda}, {Nakamura}, {Scott}, \&
  {Russell}}]{vernazza2015}
{Vernazza}, P., {Zanda}, B., {Nakamura}, T., {Scott}, E.~R.~D., \& {Russell},
  S. 2015, in Asteroids IV, 617--634

\bibitem[{{Vokrouhlick{\'y}} {et~al.}(2006){Vokrouhlick{\'y}}, {Bro{\v{z}}},
  {Morbidelli}, {Bottke}, {Nesvorn{\'y}}, {Lazzaro}, \& {Rivkin}}]{Vok2006}
{Vokrouhlick{\'y}}, D., {Bro{\v{z}}}, M., {Morbidelli}, A., {et~al.} 2006,
  \icarus, 182, 92

\bibitem[{{{\v{S}}r{\'a}mek} {et~al.}(2012){{\v{S}}r{\'a}mek}, {Milelli},
  {Ricard}, \& {Labrosse}}]{sramek2012}
{{\v{S}}r{\'a}mek}, O., {Milelli}, L., {Ricard}, Y., \& {Labrosse}, S. 2012,
  \icarus, 217, 339

\bibitem[{{Weiss} \& {Elkins-Tanton}(2013)}]{weiss2013}
{Weiss}, B.~P. \& {Elkins-Tanton}, L.~T. 2013, Annual Review of Earth and
  Planetary Sciences, 41, 529

\end{thebibliography}

%----------------------------------------------------------------- 

\begin{appendix}

\section{Spectral classes from the literature}
\label{app:lit_ref}

For this work, we used asteroid spectral classes derived from spectroscopic observations from the following references: \citet{tatsumi2022, Migliorini2021, Arredondo2021, Gartrelle2021, Hasegawa2021, depra2020, depra2020b, lucas2019, demeo2019, devogele2019, morate2019, popescu2019, perna2018, migliorini2018, depra2018, devogele2018, morate2018, Borisov2017, migliorini2017, leith2017, lucas2017, 2016Icar..266...57D, 2016A&A...586A.129M, Riberio2014, demeo2014, 2014Icar..228..217T, 2014A&A...572A.106P, Oszkiewicz2014, Sanchez2013, Gietzen2012, DeSanctis2011, deleon2010, demeo2009, clark2009, 2008Icar..195..277M, 2008Icar..194..125R, 2008A&A...487.1195L, 2008A&A...486L...9M, AlvarezCandal2006,Lazzarin2005, 2005Icar..175..170M, 2005Icar..174...54M, 2004Icar..171..120D, 2004Icar..170..259B, Lazzarin2004, bus2002tax, carvano2001}. We also used asteroid spectral classes derived from spectrophotometric observations from the following references: \cite{carvano2010,Popescu_2018_classification,SergeyevCarry2021,Sergeyev2022,demeo2013}

\section{NIR observation of (33763) 1999 RB84}
\label{app:NIR_obs}

On June 13 2023, we used the SpeX \citep{rayner2003} and MORIS instruments \citep{gulbis2011} at NASA's Infrared Telescope Facility (IRTF) to observe asteroid 33763 at 9.4$^\circ$ phase angle. SpeX was used with a slit 0.8$\times$15" in PRISM mode, covering wavelengths between 0.7 and 2.5~\SI{}{\micro\meter} with a spectral resolution of $\sim$200 in a single configuration. The asteroid's magnitude was 17.4 and the exposure time for a single frame was 180~sec. We acquired 6 cycles `AB-BA' frame pairs, for a total exposure time of 18 min, where the asteroid was shifted along the slit. Following well established procedures \citep{reddy2009}, we calculated the asteroid's reflectance $R(\lambda)$ as a function of the wavelength $\lambda$ using Eq.~\ref{E:spec}:
\begin{equation}
R(\lambda) = \frac{A(\lambda)}{S_L(\lambda)} \times 
	              Poly \left (\frac{S_L(\lambda)}{S_T(\lambda)} \right),
	              \label{E:spec}
\end{equation}
where $A(\lambda)$, $S_L(\lambda)$, $S_T(\lambda)$ are the wavelength-calibrated raw spectra respectively of the asteroid, of the local HD137782 G2V star observed within $\sim$300" of the asteroid, and of the trusted solar analogue star SA107684 that was observed at a similar airmass, respectively. The function $Poly()$ represents a polynomial fit of the stars ratio, excluding the regions affected by the telluric water vapour absorption (1.3 $< \lambda <$ 1.5, 1.78 $ < \lambda <$ 2.1, and $\lambda > $ \SI{2.4}{\micro\meter}).
Asteroid spectra were shifted to sub-pixel accuracy to align with the calibration star spectra. In general, the local star ensures accurate removal of the telluric features, but may require the slope correction 
$Poly \left (\frac{S_L(\lambda)}{S_T(\lambda)} \right)$ 
due to difference between the local star's spectrum and the one of the Sun. 

%\cite{Sanchez2013}

\section{Comparison with meteorites}
\label{app:comp_met}

We used the full VNIR spectra of asteroids 36256 and 33763 to compare these objects with the samples and meteorites available in the Relab database to gain information on their composition. To do so, we used the exponential space-weathering model derived by \cite{Brunetto2006} to correct the asteroids spectra from the effects of space weathering \citep[as done for example in][]{demeo2022,Mahlke2023}. \cite{Brunetto2006} found that the ratio between irradiated and non-irradiated silicate samples shows almost no contribution from the spectral bands, and gives a continuum that can be fitted by an exponential curve:
\begin{equation}
    W(\lambda) = K exp\left(\frac{C_s}{\lambda}\right),
    \label{eq:SW_Brunetto2006}
\end{equation}
where $\lambda$ is the wavelength, K is a normalising scale factor that depends on the normalisation of the spectra, and $C_s$ is the strength of the exponential curve, representing the strength of the space weathering. For each of the two asteroids, we calculated the ratio between their VNIR spectrum and the spectrum of every sample of the Relab database. To do so, we re-sampled the asteroids and the meteorites spectra from 450 to 2500 nm for asteroid 36256, and from 450 to 1800 nm for asteroid 33763. We adopted a narrower wavelength range for this latter asteroid, to filter out the noisier part of its NIR spectrum affected by a positive slope. Then, we fitted the exponential space-weathering model of \cite{Brunetto2006} to each calculated ratio, to compute de-weathered asteroids spectra.

We used a curve matching method to find the best analogues of asteroids 36256 and 33763, following established methods \citep[see e.g.][]{Avdellidou2022,demeo2022,Mahlke2023}. We calculated the $\Phi_{comb}$ parameter of \cite{popescu2012} between the Relab samples and the de-weathered asteroids spectra, and as the best matches, we obtained mostly olivine mineral spectra, similarly to the results from \cite{demeo2022}. Among the meteorites found as best spectral matches, we found the CK6 meteorite LEW 87009, the R-chondrite Rumuruti, the shergottite ALHA77005, the pallasites Esquel and Thiel Mountains, and the brachinite EET99402 for asteroids 36256 and 33763.

We used the equivalent geometric visible albedo presented in \cite{Beck2021} to try to discriminate between these best spectral matches. We followed the method as implemented in \cite{Avdellidou2022}: we determined the equivalent geometric visible albedo, $p_V$, of the Relab meteorite samples, to be able to compare their reflectance values to the mean albedo of the family, $p_V = 0.22$. We calculated the reflectance value at 0.55~\SI{}{\micro\meter} of the meteorites observed at a phase angle of 30$^{\circ}$, and we derived their reflectance value at a phase angle of 0$^{\circ}$, using Eq. 2 of \cite{Beck2021}. The comparison between this equivalent geometric visible albedo to the family's mean $p_V$ gives us an indication of which meteorite samples might be the best matches for these family members. Among the best meteorite analogues given by the curve matching method, we found that the CK6 meteorite LEW87009, the R-chondrites Rumuruti and LAP04840, the R4 chondrite MIL07440, the olivine of the shergottite ALHA77005, and the brachinites NWA4882 and ALH84025, show an equivalent geometric albedo close to $p_V = 0.22$. These meteorites are therefore the best matches found of asteroids 36256 and 33763. However, the shergottite ALHA77005 is a martian meteorite, and is therefore not a reasonable match for this outer-main belt asteroid family.

%shergottite ALHA77005 (R0=0.229), LAP04840 (R0=0.224), CK6 LEW87009 (R0=0.2), heated CV3 from allende (R0=0.29), R-chondrite Rumuruti (R0=0.19)

%cdts to use Beck's formula
%need g=30°
%measurements obtained on fine powders, not bulk samples
%law only applies to reflectance lower than 0.5

\section{Classification algorithm}
\label{app:algo}

We studied the possibility that the A-type classification driven by Gaia data got contaminated by objects affected by a Gaia DR3 reddening \citep{galluccio2022,galinier2023}. To investigate this issue, we tested our algorithm by classifying asteroids both having a spectrum in the DR3, and that have been classified from VIS-NIR observations by \citet{demeo2009,demeo2014,demeo2019}. This list contains 389 asteroids. We applied our classification method on their DR3 spectra, in the wavelength range from 0.45 to 0.90~\SI{}{\micro\meter}. This wavelength range was chosen in order to not take into account the first and last two bands of the DR3 spectra often affected by systematic errors \cite{galluccio2022,Oszkiewicz2023}, and to limit the effect of a potential DR3 reddening. Moreover, a quality flag going from 0 (good quality) to 2 (poorer quality) was assigned to each of the 16 bands of the Gaia DR3 spectra. We did not consider the bands flagged with a non-zero number, as done in \cite{galinier2023}.
%%%%%% i dont think it is necessary to put it.
%%% add that, to do the matching, the Bus-DeMeo templates were sampled like these cleaned Gaia DR3 spectra using a cubic smoothing spline (csaps python function) - if this is what has been done.

%How were the independent data set treated ? Were the Bus-DeMeo templates sampled like the different data sets to be able to do the matching, or were the spectra of the different data sets and the templates resampled using a cubic smoothing spline function ? Or was every data set sampled like the Bus-DeMeo templates ?

We studied the performances of our algorithm for the classification of A-types using the confusion matrix presented in Table \ref{tab:confusion_matrix}. In the list of 389 asteroids, 21 are classified as A-types. We found that eight asteroids that are not A-type in \citet{demeo2009,demeo2014,demeo2019} are classified A-type by our method. This corresponds to a false positive rate of around 2.2\%. However, among these false positives, three asteroids were later classified as A-type by \cite{Mahlke2022} in his own taxonomic scheme. If we consider this information, the false positive rate goes down to less than 1.5\%.

Among the 21 A-type asteroids of the list, 15 get correctly classified as A-types as first or second best match with the Bus-DeMeo templates. This corresponds to a true positive rate of above 71\%. The accuracy of the classification corresponds to the sum of the true positive and the true negatives, divided by the total population. The accuracy for the A-type classification here is therefore of $\frac{15+360}{389}=96$\%.

%%% to be checked with the right algorithm, might be different since the error bars are taken into account here !
Among the six A-type asteroids of the literature that did not get classified as A-type here (false negative rate of 28.6\%), four asteroids have a S/N$<$21. We conclude that the classification of low S/N A-type asteroids should be taken with more caution: it is not impossible that a low S/N asteroid that was not found A by our method is an A. Similarly, asteroids found A by our method and that show a S/N<21 might not be real A.

%%% reduced Chi2 close to 1 : good fit. Chi2>1 can be improved, and Chi2<1 overfit. So that's why be careful with the S/N, risk of overfitting

We did not aim to extensively study the performance of our classification algorithm on other taxonomic types. We therefore advise caution with the use of spectral matching between DR3 spectra and Bus-DeMeo templates to classify asteroids in the DR3 dataset.

\begin{table}\centering %\setlength\tabcolsep{3.5pt}\renewcommand\arraystretch{1.25}
\caption{Confusion matrix of the classification of A-type asteroids by our algorithm. The list of 389 asteroids both classified by \citet{demeo2009,demeo2014,demeo2019} and having a DR3 spectrum considered for this study contains 21 A-type asteroids.}
    \begin{tabular}{|l|*{2}{c|}}
      \hline
      \diagbox[width=\dimexpr \textwidth/8+2\tabcolsep\relax, height=1cm]{Predicted}{True}
      & A-type & Non-A-type \\
      \hline
      A-type & 15 & 8 \\
      \hline
      Non-A-type & 6 & 360 \\
      \hline
    \end{tabular}
  \label{tab:confusion_matrix}
\end{table}

\newpage

\section{Figures}

\begin{figure*}
\includegraphics[width=0.3\textwidth]{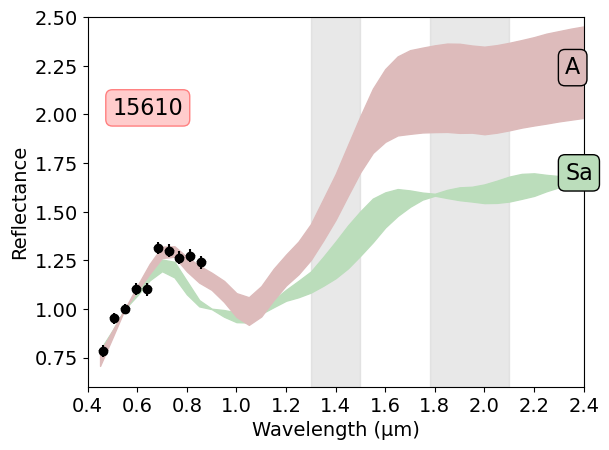} 
\includegraphics[width=0.3\textwidth]{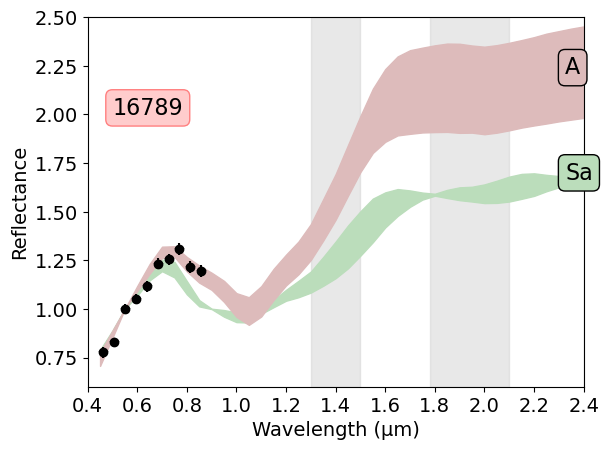} 
\includegraphics[width=0.3\textwidth]{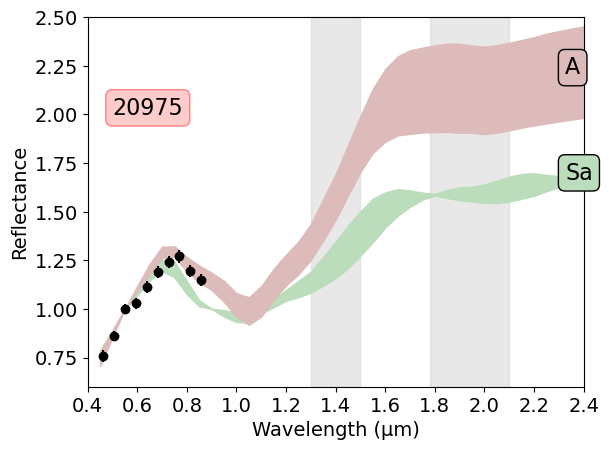}\\ 
\includegraphics[width=0.3\textwidth]{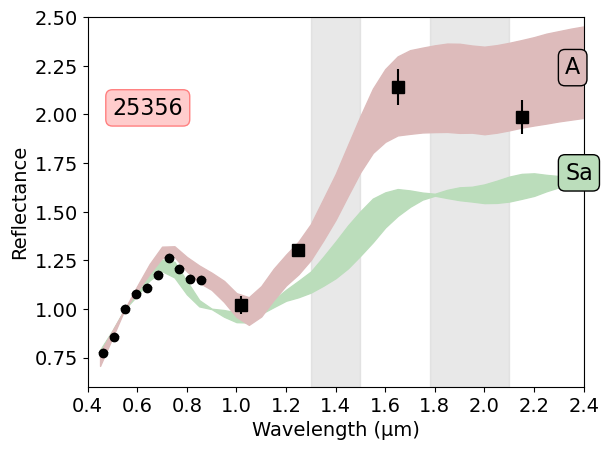} 
\includegraphics[width=0.3\textwidth]{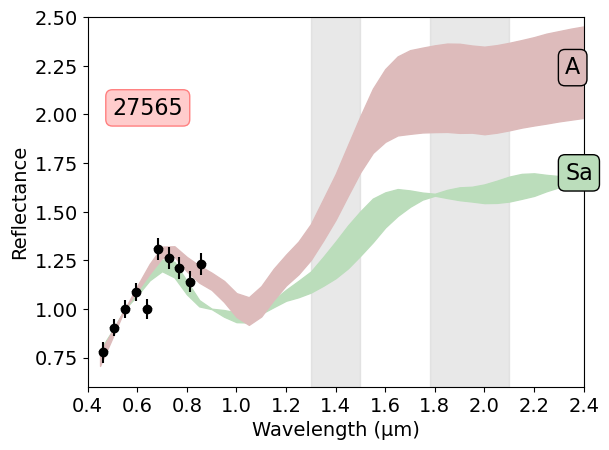} 
\includegraphics[width=0.3\textwidth]{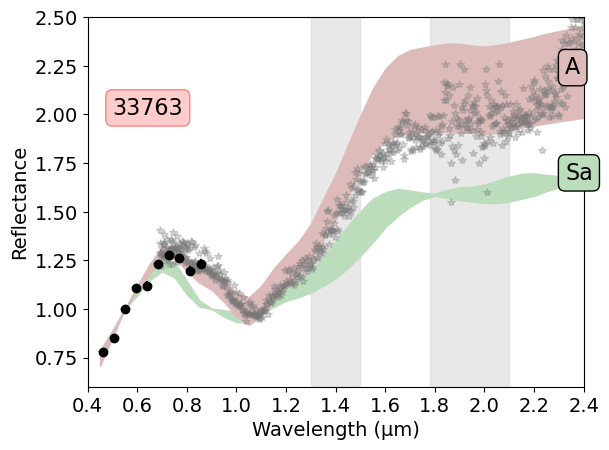} \\
\includegraphics[width=0.3\textwidth]{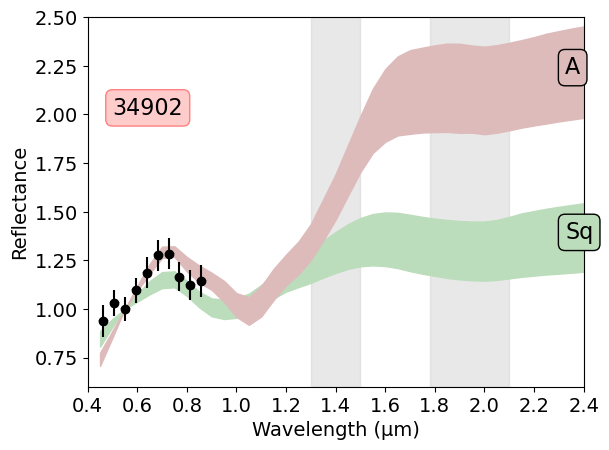} 
\includegraphics[width=0.3\textwidth]{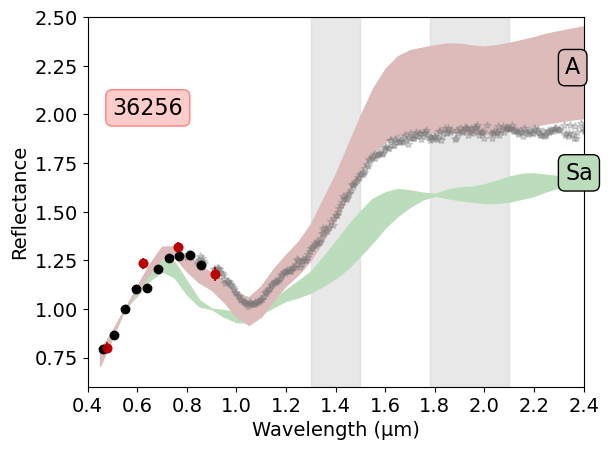}
\includegraphics[width=0.3\textwidth]{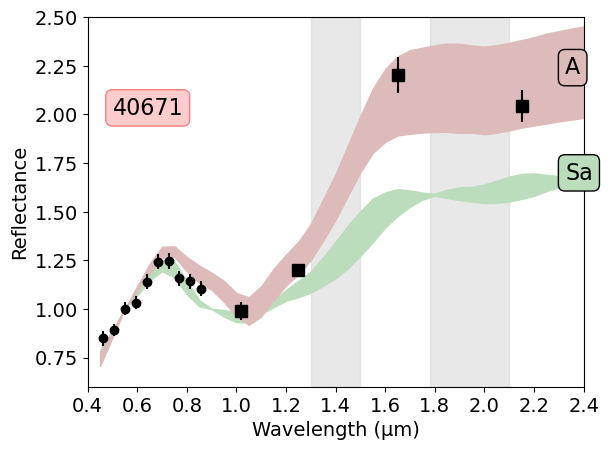} \\
\includegraphics[width=0.3\textwidth]{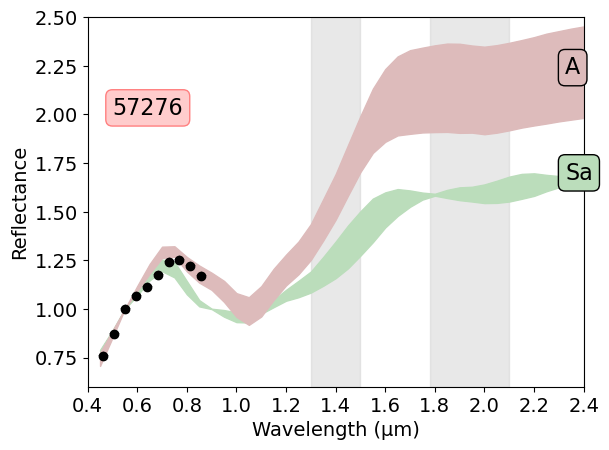} 
\includegraphics[width=0.3\textwidth]{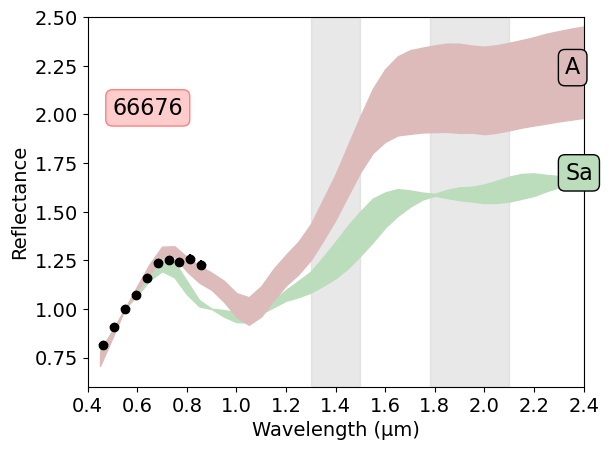} 
\includegraphics[width=0.3\textwidth]{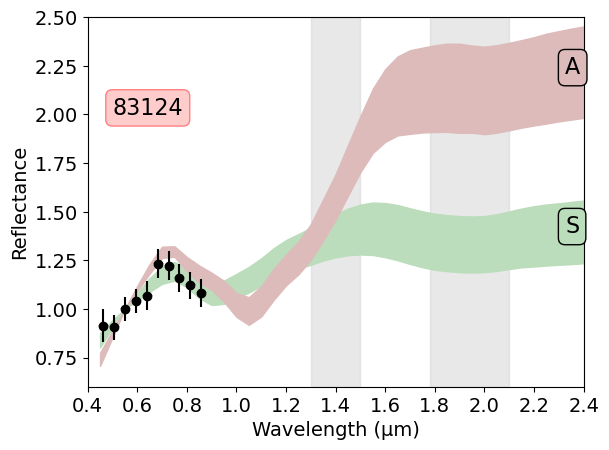} \\
\includegraphics[width=0.3\textwidth]{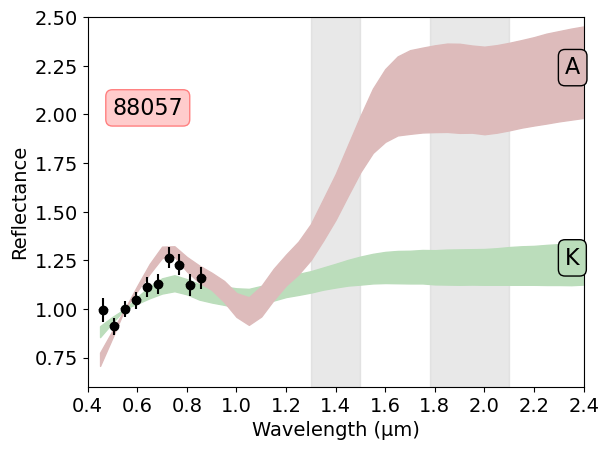} 
\includegraphics[width=0.3\textwidth]{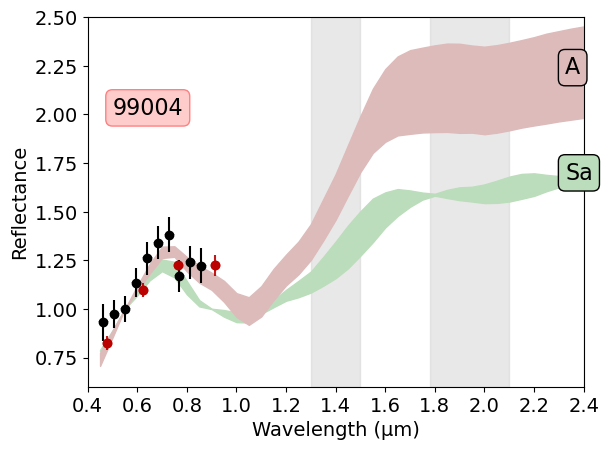} 
\includegraphics[width=0.3\textwidth]{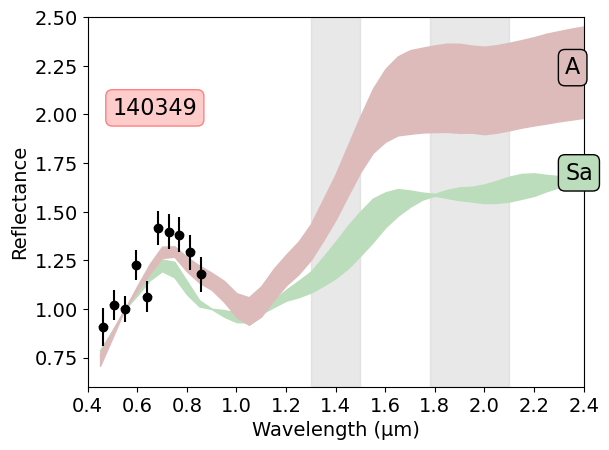} 
\caption{Available data of the members of the family (36256) 1999~XT17 in the Gaia DR3: Gaia spectra (black dots), SDSS data from \cite{demeo2013} (red dots), MOVIS data from \cite{Popescu_2018_classification} (black squares), and NIR spectra from \cite{demeo2019} (grey stars). The shaded coloured areas represent the reflectance spectra from the A-type and other classes templates of \cite{demeo2009}. \label{fig:all_spectra}}
\end{figure*}

\begin{figure*}[!ht]
    \centering
    \begin{subfigure}{0.5\textwidth}
        \includegraphics[width=\textwidth]{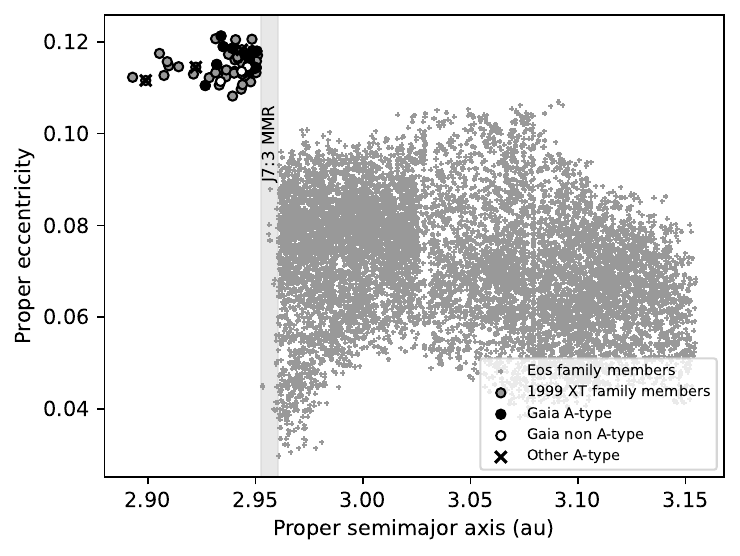}
        \caption{}
        \label{fig:subfiga}
    \end{subfigure}%
    \begin{subfigure}{0.5\textwidth}
        \includegraphics[width=\textwidth]{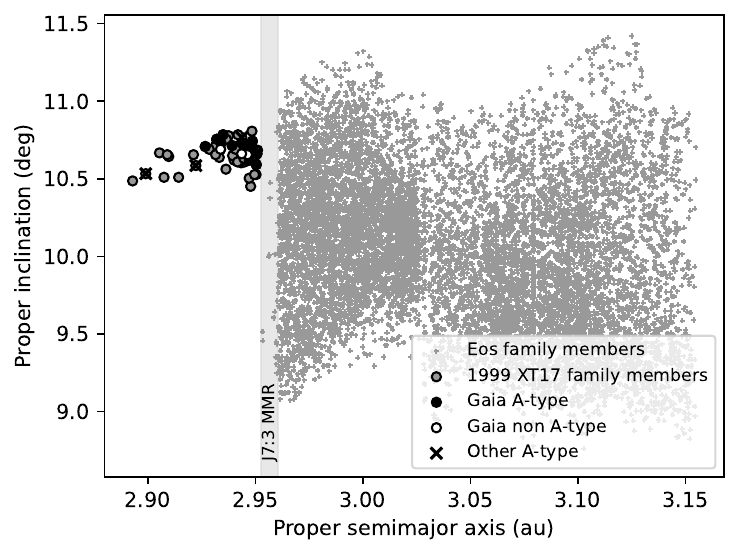}
        \caption{}
        \label{fig:subfigb}
    \end{subfigure}
    \begin{subfigure}{0.5\textwidth}
        \includegraphics[width=\textwidth]{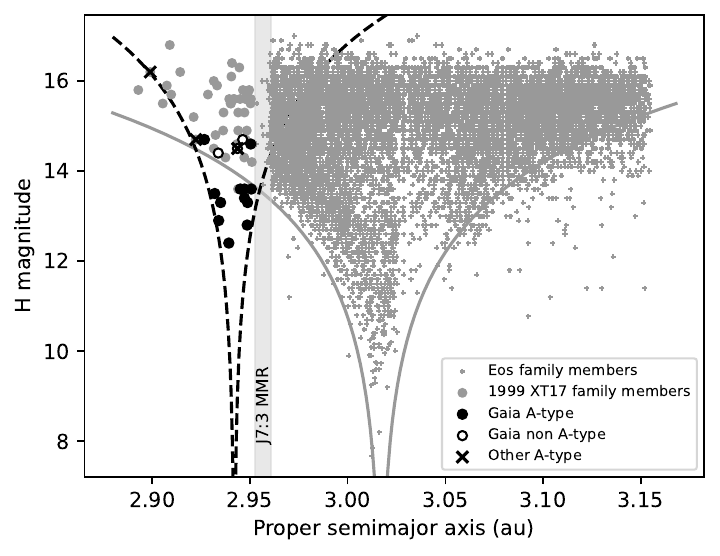}
        \caption{}
        \label{fig:subfigc}
    \end{subfigure}
    \caption{Proper orbital elements (panel a and b) and V-shape (panel c) plots of the 1999~XT17 family and the (221) Eos family. %We used family memberships as defined by \cite{nesvorny2015}. 
The members of Eos are displayed as grey crosses, and those of the 1999~XT17 family as circles. Black filled circles are 1999~XT17 family members that were classified as A-types by our algorithm, while the empty circles are members that were not classified as A-types. Family members that were classified A-types in the literature and have no DR3 spectra are marked as "X". Among the latter group, asteroid 76627 overlaps in the plot with asteroid 88052 (non A-type), due to their similar proper orbital elements and H magnitudes. The position of the 7:3 mean motion resonance with Jupiter is highlighted with a shade of light-grey. The V-shapes of the two families are drawn by eye following the equation $H = 5 \log_{10} (|a - a_c|/C_0)$ with $a_C=2.942$ au and $C_0=1.8\times10^{-5}$ au for the 1999~XT17 family, and $a_C=3.017$ au and $C_0=2.25\times10^{-4}$ au for the Eos family.}
    \label{fig:Vshape}
\end{figure*}

\end{appendix}

\end{document}